\documentclass[aps,reprint,showpacs,twocolumn,tightenlines]{revtex4-2}
\usepackage{amsmath,amssymb,graphicx}
\usepackage{color}
\usepackage{ulem}
\usepackage{lineno}
\usepackage{dcolumn}
\usepackage{braket}
\usepackage{amssymb}
\usepackage{amsmath}
\usepackage{latexsym}
\usepackage{mathrsfs}
\usepackage{arydshln}
\usepackage[sans]{dsfont}
\usepackage{epsfig}
\usepackage{epstopdf}
\usepackage{natbib}
\usepackage{lipsum}
\usepackage{amsthm,bm}

\newcommand{\bitem}{\begin{itemize}}
	\newcommand{\fitem}{\end{itemize}}
\newcommand{\beq}{\begin{equation}}
	\newcommand{\eeq}{\end{equation}}
\newcommand{\beqa}{\begin{eqnarray}}
	\newcommand{\eeqa}{\end{eqnarray}}

\newcommand{\blue}[1]{{\color{black} #1}}

\begin{document}

\title{Quantum Neuronal Sensing of Quantum Many-Body States on a 61-Qubit Programmable Superconducting  Processor}

\author{Ming Gong$^{1,2,3}$}
\thanks{These authors contributed equally to this work.}
\author{He-Liang Huang$^{1,2,3}$}
\thanks{These authors contributed equally to this work.}
\author{Shiyu Wang$^{1,2,3}$}
\thanks{These authors contributed equally to this work.}
\author{Chu Guo$^{1,2,3}$}
\author{Shaowei Li$^{1,2,3}$}
\author{Yulin Wu$^{1,2,3}$}
\author{Qingling Zhu$^{1,2,3}$}
\author{Youwei Zhao$^{1,2,3}$}
\author{Shaojun Guo$^{1,2,3}$}
\author{Haoran Qian$^{1,2,3}$}
\author{Yangsen Ye$^{1,2,3}$}
\author{Chen Zha$^{1,2,3}$}
\author{Fusheng Chen$^{1,2,3}$}
\author{Chong Ying$^{1,2,3}$}
\author{Jiale Yu$^{1,2,3}$}
\author{Daojin Fan$^{1,2,3}$}
\author{Dachao Wu$^{1,2,3}$}
\author{Hong Su$^{1,2,3}$}
\author{Hui Deng$^{1,2,3}$}
\author{Hao Rong$^{1,2,3}$}
\author{Kaili Zhang$^{1,2,3}$}
\author{Sirui Cao$^{1,2,3}$}
\author{Jin Lin$^{1,2,3}$}
\author{Yu Xu$^{1,2,3}$}
\author{Lihua Sun$^{1,2,3}$}
\author{Cheng Guo$^{1,2,3}$}
\author{Na Li$^{1,2,3}$}
\author{Futian Liang$^{1,2,3}$}
\author{Akitada Sakurai$^{4,6}$}
\author{Kae Nemoto$^{6,7,4}$}
\author{W. J. Munro$^{5,6}$} \email{bill.munro@me.com}
\author{Yong-Heng Huo$^{1,2,3}$}
\author{Chao-Yang Lu$^{1,2,3}$}
\author{Cheng-Zhi Peng$^{1,2,3}$}
\author{Xiaobo Zhu$^{1,2,3}$} \email{xbzhu16@ustc.edu.cn}
\author{Jian-Wei Pan$^{1,2,3}$} \email{pan@ustc.edu.cn}

\affiliation{$^1$ Department of Modern Physics, University of Science and Technology of China, Hefei 230026, China}
\affiliation{$^2$  Shanghai Branch, CAS Center for Excellence in Quantum Information and Quantum Physics, University of Science and Technology of China, Shanghai 201315, China}
\affiliation{$^3$  Shanghai Research Center for Quantum Sciences, Shanghai 201315, China}
\affiliation{$^4$  School of Multidisciplinary Science, Department of Informatics, SOKENDAI (the Graduate University for Advanced Studies), 2-1-2 Hitotsubashi, Chiyoda-ku, Tokyo 101-8430, Japan}
\affiliation{$^5$  NTT Basic Research Laboratories and Research Center for Theoretical Quantum Physics, 3-1 Morinosato-Wakamiya, Atsugi, Kanagawa 243-0198, Japan}
\affiliation{$^6$  National Institute of Informatics, 2-1-2 Hitotsubashi, Chiyoda-ku, Tokyo 101-8430, Japan}
\affiliation{$^7$ Okinawa Institute of Science and Technology Graduate University, Onna-son, Okinawa 904-0495, Japan}

\begin{abstract}
Classifying many-body quantum states with distinct properties and phases of matter is one of the most fundamental tasks in quantum many-body physics. However, due to the exponential complexity that emerges from the enormous numbers of interacting particles, classifying large-scale quantum states has been extremely challenging for classical approaches.
Here, we propose a new approach called quantum neuronal sensing. Utilizing a 61 qubit superconducting quantum processor, we show that our scheme can efficiently classify two different types of many-body phenomena: namely the ergodic and localized phases of matter. Our quantum neuronal sensing process allows us to extract the necessary information coming from the statistical characteristics of the eigenspectrum to distinguish these phases of matter by measuring only one qubit. Our work demonstrates the feasibility and scalability of quantum neuronal sensing for near-term quantum processors and opens new avenues for exploring quantum many-body phenomena in larger-scale systems.
\end{abstract}

\maketitle


It is well known that one of the exciting applications for quantum processors is associated with the probing and characterization of complex phenomena in quantum many-body systems~\cite{coleman2015introduction,bernien2017probing, zhang2017observation, ye2019propagation, chen2021observation, gong2021experimental,ebadi2021quantum,eisert2015quantum,schweigler2017experimental,lanyon2017efficient,zhao2020quantum,thomas2018experimental,prufer2020experimental,guo2021observation}. Quantum many-body systems are known to exhibit various fascinating phases of matter, including high-temperature superconductivity, spin liquids, and quantum Hall effects to mention but a few~\cite{savary2016quantum,wen1992theory}. However, both theoretical and experimental approaches to explore such systems suffer severe limitations due to the complex and high dimensional nature of these systems \cite{felser2021efficient,melko2019restricted,carleo2018constructing,vicentini2021machine}. By contrast, quantum processors are expected to be able to generate these interesting phases of matter in a highly controllable and coherent manner \cite{bernien2017probing,zhang2017observation,gong2021quantum,ebadi2021quantum}. The size of the Hilbert space associated with a 60+ qubit quantum processor~\cite{gong2021quantum,wu2021strong,zhu2021quantum} should be sufficient to explore various quantum statistical properties of such phases of matter without resorting to the usual approximations \cite{gao2017efficient,cai2018approximating,bravyi2019approximation}. This should allow us to explore their properties beyond the understanding conventional techniques provide \cite{carleo2017solving,carrasquilla2017machine,vidal2004efficient}. However, using these powerful noisy intermediate-scale quantum (NISQ) processors brings several serious complications associated with their use - beyond those associated with noise and imperfections. It is well-known that the amount of the information that can be extracted from quantum states is limited, and although we can reconstruct the quantum state or process through tomographic techniques \cite{lanyon2017efficient}, the number of measurements usually grows exponentially with the processor size ~\cite{cramer2010efficient,torlai2018neural}. This could make the entire effort for many-body dynamics simulation on a quantum information processor inefficient.~
To overcome these limitations and enable efficient quantum simulation of such systems, we need {not} only an appropriate means to generate our required dynamics, but more importantly, a measurement scheme to observe it that scales favorably with the processor size.

In this work, we propose a quantum neuronal sensing approach to determine the statistical properties associated with the states of quantum many-body systems. It combines quantum sensing in which one observes the fluctuation of a system parameter with the computational power of quantum neural networks (QNNs).
{The usual quantum information approaches \cite{Orell2019} to explore quantum many-body physics and confirm their properties, such as the particular quantum phase of matter, tend to be computationally intensive. Further such approaches tend to be rather sensitive to finite system size effects.}
Instead, we consider the state generated in the many-body system as a quantum input to the sensor where a QNN is used to preprocess the state for the targeted information to be efficiently extracted onto just one qubit. That qubit is then measured.
{As the input data to the QNN is a quantum state, and the QNN can directly process the information encoded in that quantum state, it has a natural advantage against its classical counterparts.  If the QNN can help the readout extract more information about the knowledge we aim to have, this quantum neuronal sensing is positive in extending its sensing abilities.}
Our quantum neuronal sensing approach could have a quantum advantage for both the direct quantum information processing before the measurement and the use of the quantum computational power of the process. We demonstrate our quantum sensing scheme using a NISQ-era 61-qubit programable superconducting quantum processor. This is first used to generate nontrivial quantum statistical properties, including the quantum ergodic and localized phases of matter. Then, quantum neuronal sensing probes the system to determine whether it is in an ergodic or localized phase by measuring only one qubit.

\begin{figure}
\centering 
\includegraphics[width=0.9\linewidth]{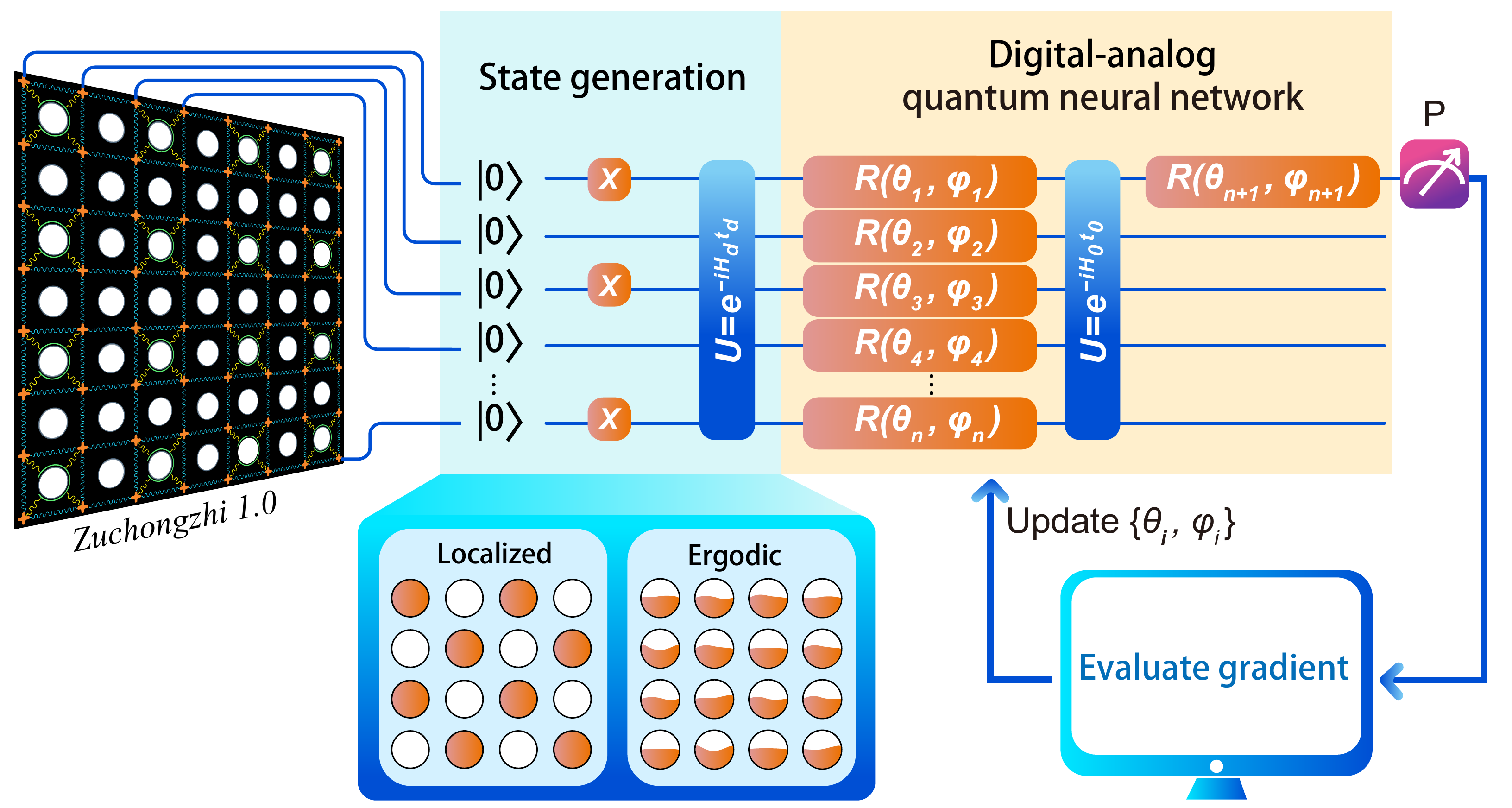}
\caption{\textbf{The scheme for quantum neuronal sensing of quantum many-body states using QNNs.} The experiment is performed on a superconducting quantum processor \textit{Zuchongzhi 1.0}. It begins with the state preparation, where we generate either the localized or ergodic state. The state is the input state of the QNN, which in our case is a double-layered digital-analog variational quantum circuit. Here we use $N_q+1$ trainable single-qubit rotations $R(\theta_i,\phi_i)$ and a multi-qubit unitary operation $U$ as the ansatz. A quantum-classical hybrid architecture implements the training of the QNN. The classical computer is used to evaluate gradient from the measurement results marked as ``P'', which is the probability of measuring the readout qubit in its $\ket{1}$ state. After the training, the QNN can classify the states in localization or ergodicity with the readout of only one qubit.}
\label{fig1}
\end{figure}

At the core of our quantum neuronal sensing scheme is a digital-analog QNN whose task is to distinguish the quantum states, namely ergodic and localized states in this case.
Our experiments generate these two distinct groups of many-body states by a superconducting quantum processor composed of 64 qubits in a square $8\times8$ lattice arrangement. Among them, two qubits are not functional while another is not fast tunable~(see Supplementary Materials~\cite{seesm}), reducing the maximum number of operational qubits to 61. This scheme requires the preparation of training states which we achieve in the following way. All qubits in our processor are first initialized in their ground $\ket{0}$ states and then prepared to the Neel state by exciting every alternate qubit. Our system then evolves under the interaction Hamiltonian $H_d=\hbar \sum g_{i,j}(\sigma_x^i\sigma_x^j+\sigma_y^i\sigma_y^j)/2+ \hbar \sum d_i \sigma_z^i$ for a time $t_d=200$ ns where $g_{i,j}$ is the effective coupling strength between neighboring qubits with the average coupling for all neighboring qubit sets given by $g /2 \pi=2.75$ MHz. Further $d_i$ is the disorder for qubit $Q_i$ randomly selected from a uniform distribution $[-h, h]$ with {$h_{\text{erg}}/2 \pi=1$ MHz ($h_{\text{loc}}/2 \pi=50$ MHz) chosen} for the generation of ergodic (localized) states respectively. \blue{We note that while the existence of exact two-dimensional localization is still debated, in a finite-size system, localized states are observed in both numerical analysis~\cite{Kuhn_2007} and experiential demonstration~\cite{Escalante2018,Zhang2021}. As shown in the Supplementary Materials Fig.S7~\cite{seesm}, the dynamics of the time evolutions of the system imbalance at different disorder strengths signal evidence of the systems entrance into the localized phase under strongly disordered two-dimensional systems, which are observed for the first time in a large-scale programmable superconducting quantum processor.}

Now that we have generated our quantum states, we can directly process and analyze them using a QNN based on a variational quantum circuit~\cite{cong2019quantum,liuqccnn2021,benedetti2019parameterized} built in a hybrid digital-analog architecture.
The variational circuit consists of $N_q+1$ single qubit rotations $R(\theta_i, \phi_i)=Z(\theta_i)X(\phi_i)Z(-\theta_i)$ {(digital part)} and a multi-qubit unitary operation $U=e^{-i H_0 t_0/ \hbar}$ (analog part) in the two layers. In the first layer the single qubit rotations are parameterized by the trainable parameters $\theta$ and $\phi$ while for the analog circuit $H_0=\sum \hbar g(\sigma_x^i\sigma_x^j+\sigma_y^i\sigma_y^j)/2$ is the $N_q$-qubit system Hamiltonian with zero disorder on each site. The analog circuit operates for a time $t_0=200$ ns. Next the second layer of the QNN contains only one single-qubit rotation for the measured qubit (see Fig.~\ref{fig1}B). This means the total number of parameters in this configuration for $N_q$ qubits is $2(N_q+1)$, which can be extended, if necessary, to $N_l+1$ layers with $2(N_q N_l+1)$ parameters.

The natural question is how our quantum neuronal sensing scheme works, noting that the analog circuit part quickly generates entanglement while the digital elements are utilized for training the expression of the circuit. The features of the input data are expected to be extracted through the trained variational quantum circuit and then mapped to a state that is easy to recognize with deterministic measurement outcomes. Here that binary classification is achieved by measuring only one qubit. The input state is recognized as a localized state if the probability of measuring the readout qubit in its $|1\rangle$ state is smaller than 0.5; otherwise, we consider it an ergodic state. \blue{The threshold of 0.5 is the default value for the binary classification problem, if the normalized predicted probabilities are in the range between 0 and 1. However, the threshold does not have to be 0.5, and it can easily be adjusted to the optimal value if a class imbalance appears after the neural network.} 

Next, the training of the QNN is accomplished by a gradient-based method generally preferred with large parameter spaces \cite{mitarai2018quantum}. Given we are trying to determine if the state is ergodic or localized, we use the binary cross-entropy loss function~\cite{ruby2020binary,mannor2005cross,ho2019real}
\begin{eqnarray}
L({\bm{\theta,\phi}}) &=& \frac{1}{N}\sum\limits_i - [({y_i}\cdot {w_i} \cdot \log ({p_i})\nonumber \\
&\;&\;\;\;\;\;\;\;\;\;\;\;+ (1 - {y_i}) \cdot {w_i} \cdot \log (1 - {p_i})]
\label{loss}
\end{eqnarray}
to optimize the single-qubits trainable parameters ${(\bm{\theta, \phi})}$, where $y_i$ is the label of the $i$-th training data with $y_i=1$ if the state is ergodic and 0 otherwise. Next $w_i$ is the weight of $i$-th data with it set as 3 (1) for the ergodic (localized) states respectively for a better training performance. Further $p_i$ and $1-p_i$ are the probabilities for the class of ergodic states and localized states, respectively. The gradient of the loss function (\ref{loss}) can be evaluated by the parameter-shift rules~\cite{mitarai2018quantum,schuld2019evaluating}
\begin{eqnarray}
\frac{{\partial L(\bm{\theta ,\phi} )}}{{\partial {\theta _j}}} &=&
\frac{1}{2}\{ L[(\bm{\theta ,\phi} ){\rm{ + (0,}}...{\rm{,}}\frac{\pi }{2}{\rm{,}}...{\rm{0)]}} \nonumber \\
&\;&\;\;\;\;\;\;\;\;\;\;\;- L[(\bm{\theta ,\phi} ) - {\rm{(0,}}...{\rm{,}}\frac{\pi }{2}{\rm{,}}...{\rm{0)]}}\}
\label{shiftrule}
\end{eqnarray}
where the $\pi /2$ factor appears at the $j$-th position. 

\begin{figure*}
\centering \includegraphics[width=0.8\linewidth]{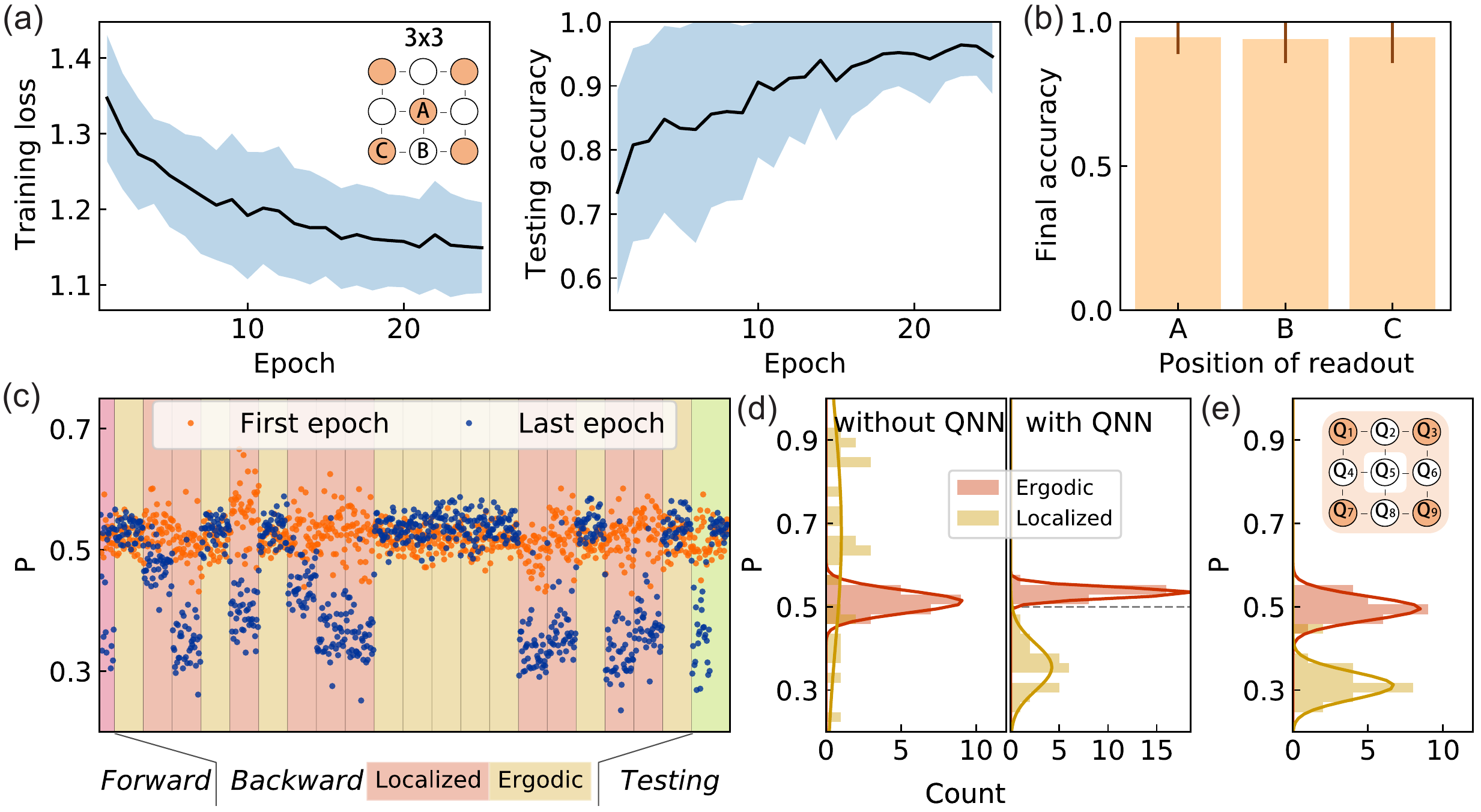}
\caption{\textbf{The training and testing of the QNN for qubits arranged in $3\times3$ grid.} (a) The loss and accuracy at every epoch during the training. 
The solid black lines are averaged over ten independent training instances, while the blue shadows are the corresponding standard deviations. The inset is the arrangement of the nine qubits used in this case. The orange-filled circles are the qubits excited to $\ket{1}$ state in state input. The qubit in the center is the qubit for readout. \blue{(b) The testing accuracy of the readout qubit in the center (A), in the edge (B), and in the corner (C). The 
average accuracies are 0.946$\pm$0.058, 0.940$\pm$0.083, and 0.947$\pm$0.089, respectively.} (c) The intermediate data for the first and last epochs in a complete training {cycle}, where the ordinate represents the probability {$P$} of measuring {the single qubit in the $|1\rangle$ state}. 
Each epoch can be divided into three stages, \textit{Forward}, \textit{Backward}, and \textit{Testing}. The \textit{Forward} and \textit{Testing} stages can be understood as testing the QNN on the training and testing sets, respectively. In the \textit{Backward} stage, for each quantum state in the training data set (20 quantum states), we use the parameter-shift rule to train the 20 trainable parameters in the neural network. For each quantum state, a total of 40 different circuits need to be implemented and measured to calculate the gradient of those 20 training parameters. It was found that the results of the first epoch (orange dots) are not distinguishable, while those of the last epoch (blue dots) are clearly distinguishable. \blue{(d) The results of the readout qubit in the center with and without the trained QNN.} The separation of the two different groups is clearly enhanced after 25 epochs of training. The solid lines represent Gaussian fitting to the distributions of data. \blue{(e) The classification of states using the QNN where the probe qubit is not involved in state preparation. The ergodic and localized states can be classified with an accuracy 0.98. The inset is the layout of the qubits, where the qubit in the center Q$_5$ is not involved in state preparation while being used as the probe qubit. }}
\label{fig2}
\end{figure*}

We need to verify the effectiveness of our quantum neuronal sensing scheme on a small-scale quantum processor to enable us to check the results using classical simulations. As such, we start with the simplest case with nine qubits arranged in $3\times3$ grid as depicted in Fig.\ref{fig2}(a) where we use the middle qubit, marked as A, for readout. We train the QNN for several epochs until it converges, where an epoch is one complete pass through the training data. At each epoch, 
there are 20 (50) individual states in the training (testing) data, respectively, of which half are ergodic states and the remaining localized states. 
In Fig.\ref{fig2}(a) we plot the loss and accuracy values of 25 epochs during the training of the QNN, where each point is the average of 10 independent training of the QNN with initial parameters and dataset randomly generated. We can observe that the loss function gradually decreases to convergence. Meanwhile, the accuracy of the QNN on the testing set is improved synchronously. The accuracy averaged over ten different training instances converges to greater than 0.90 in 10 epochs and finally approaching 0.95. \blue{In Fig.~\ref{fig2}(b), we also provide the results for the QNN with another position as the readout qubit. In all these cases, a final classification accuracy close to 0.95 was achieved, showing that there is no specificity for different readout positions.} To investigate the details during the optimization, we present intermediate data for the first and last epochs in a complete training cycle (see Fig.~\ref{fig2}(c)). A distinct improvement of the distinguishability is observed. \blue{A more intuitive comparison is provided in Fig.~\ref{fig2}(d), which shows the measurement results of the readout qubit to the testing set with and without applying the trained QNN. The testing data are clearly distinguished into two groups after applying the trained QNN, whereas without QNN there is a large overlap between the measurement results of the two classes of states. To further verify the effect of QNN, we specially design an experiment, in which we set the qubits Q$_1$-Q$_4$ and Q$_6$-Q$_9$ as system qubits, and Q$_5$ as a probe qubit. Q$_5$ is not involved in the preparation of many-body quantum state in the state generation stage, however, it participates in the process of QNN and acts as the readout qubit. In this case, after training the QNN, measuring Q$_5$ can still clearly distinguish the two classes of quantum states (see Fig~\ref{fig2}(e) with further details provided in Supplementary Materials~\cite{seesm}). This series of experiments clearly shows that QNN act as quantum-enhanced data processor for classifying many-body quantum states,} and such high-quality training performance lays down a solid basis for further applications.

\begin{figure*}[hbt]
\centering \includegraphics[width=0.8\linewidth]{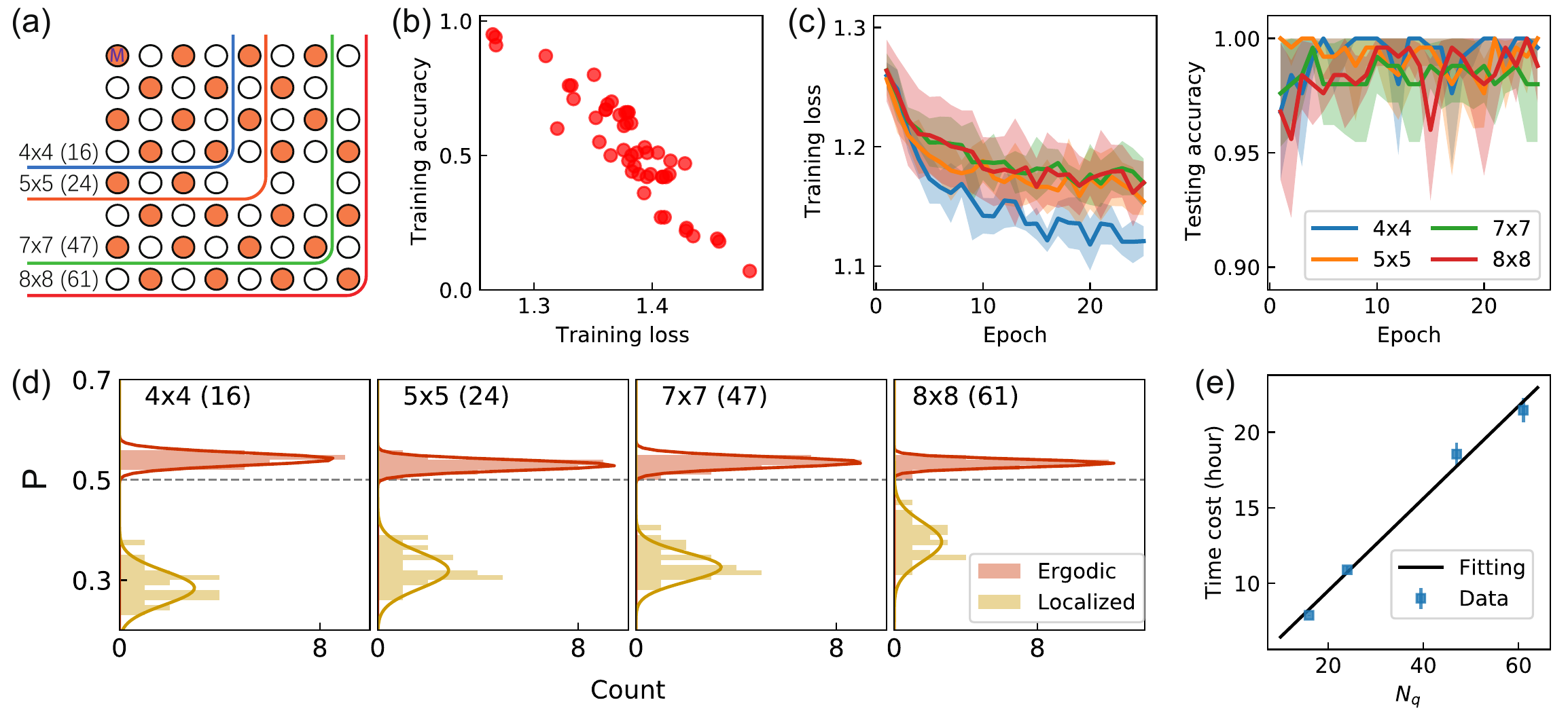}
\caption{\textbf{The training and testing of the QNN for larger-scale systems.} (a) The active qubit arrangement used in our quantum processor to realize four different system sizes ranging from $4\times 4$ to $8\times 8$, with the number of qubits (shown in the brackets) increasing from 16 to 61. The qubits excited to $\ket{1}$ state in state generation are represented as orange-filled circles. 
(b) The relationship between the loss function and the accuracy for the $8\times 8$ case. Before the training of the QNN, we randomly sampled 50 sets of parameters to choose the best initial parameter set with the lowest loss. (c) The loss and accuracy as a function of the epoch during the training. The blue, orange, green, and red solid lines are averaged over five independent instances for the four system sizes. The corresponding shadow is the error estimated via the standard deviation in different instances. (d) The distribution of testing data after 25 epochs of training. In all four cases, the results for ergodic and localized states are well-separated into two different groups. The solid lines are the Gaussian fittings to the distributions of data. (e) The training time in hours as a function of qubit number $N_q$. The solid black line is the linear fitting to the data.}
\label{fig3}
\end{figure*}

The small-scale $3\times3$ qubit processor was able to demonstrate the effectiveness of our quantum neuronal sensing scheme for distinguishing ergodic and localized states while being simple enough to be classically simulated. 
An increased number of parameters with a larger $N_q$ indicates the difficulty of the QNN optimization. However, the simplicity of the second layer of QNN helps the optimization to be feasible. In general, the training of the neural network is highly interdependent with the initial parameters. Thus, a parameter initialization strategy is employed to search for a good starting point. To achieve this good initial parameter set, we randomly generated 50 sets of parameters. Then we used a quantum data set containing 100 individual states (50 ergodic and 50 localized states) to establish the value of loss function and classification accuracy for each parameter set. In the 61 qubit instance arranged in $8\times8$ grid (Fig.\ref{fig3}(a)), we show the relationship between the accuracy and the loss function with the 50 randomly generated sets of parameters in Fig.\ref{fig3}(b). It clearly shows the trajectory that the lower loss function is associated with higher accuracy. Moreover, with only 50 randomly generated parameter sets, three sets of parameters allow the QNN to reach an accuracy greater than 0.9, showing the high availability of suitable parameters provided by our initialization strategy.

Now starting with the best initial parameters previously established, we finished our training procedures with the decreasing of the loss function and the increasing of accuracy for different system size: $4\times4$, $5\times5$, $7\times7$, and $8\times8$ (see Fig.\ref{fig3}(a)). Fig.\ref{fig3}(c) shows each curve averaged over five individual training instances. It is essential to mention here that although we start with the best initial parameters with relatively low loss function already, our training procedure can still reduce the loss function {further enabling increased separation of the ergodic and localized distributions}. As shown in Fig.\ref{fig3}(d), the trained QNN at each system size can accurately predict the previously unseen quantum state in the testing set, by measuring only one qubit, instead of measuring the entire quantum state (or populations) as in conventional methods. Our protocol avoids the exponential cost of quantum state readout, making it more valuable and scalable for large-scale quantum systems. We might expect in ideal conditions that increasing the system size can help the state classification, but Fig.\ref{fig3}(d) shows a slight decrease which we attribute to noise and imperfections in our processor's operations, which goes with system size. Regardless, our ability to predict properties of unseen quantum states also shows the potential of our scheme to assist us in detecting and characterizing different phases of matter. Finally, we also found that the time cost for training increases linearly in system size (Fig.\ref{fig3}(e)), with the largest 61 qubit system taking 22 hours, indicating the scalability of our protocol for larger systems. \blue{Moreover, the trained QNN can be applied to a wider set of unseen data (shown in the Supplementary Materials~\cite{seesm}), i.e, the quantum systems with different disorders and evolution times compared to the training set, showing the strong generalization of QNN.}

Our successful demonstration of quantum neuronal sensing of the quantum many-body system using one measurement qubit is an essential step forward for NISQ era near-term applications. Our QNN based scheme can learn and extract the highly relevant features from the information hidden in the exponentially high-dimensional Hilbert space by directly processing the information of the input state without measurement.
{The quantum neural sensing could bypass the fundamental limit of the extractable information from a quantum state, which would be a huge advantage in the future quantum sensing design.}
The digital-analog quantum framework we used for the simulation of many-body dynamics is hardware-efficient and also offers universality for quantum computing~\cite{ParraDigital2020}. {The simplicity and efficiency of our approach can naturally be extended to explore a wide range of fascinating quantum phenomena, as well as practical quantum computing, such as probing the complex phase diagram \cite{cong2019quantum}, identifying the entanglement of different structures, and been designed as a decoder for quantum error-correcting codes.}

\begin{acknowledgments}
The authors thank the USTC Center for Micro- and Nanoscale Research and Fabrication for supporting the sample fabrication and Victor M. Bastidas for valuable discussions.
The authors also thank QuantumCTek Co., Ltd., for supporting the fabrication and the maintenance of room-temperature electronics.
This research was supported by the National Key R\&D Program of China, Grant 2017YFA0304300, the Chinese Academy of Sciences, Anhui Initiative in Quantum Information Technologies, Technology Committee of Shanghai Municipality, National Science Foundation of China (Grants No. 11905217, No. 11774326),  Natural Science Foundation of Shanghai (Grant No. 19ZR1462700), and Key-Area Research and Development Program of Guangdong Provice (Grant No.2020B0303030001). This work was also supported in part by the Japanese MEXT Quantum Leap Flagship Program (MEXT Q-LEAP), Grant No. JPMXS0118069605. H.-L. H. acknowledges support from the Youth Talent Lifting Project (Grant No. 2020-JCJQ-QT-030), National Natural Science Foundation of China (Grants No. 11905294, 12274464), China Postdoctoral Science Foundation, and the Open Research Fund from State Key Laboratory of High Performance Computing of China (Grant No. 201901-01).
\end{acknowledgments}
	
\bibliography{references}

\begin{thebibliography}{59}%
\makeatletter
\providecommand \@ifxundefined [1]{%
 \@ifx{#1\undefined}
}%
\providecommand \@ifnum [1]{%
 \ifnum #1\expandafter \@firstoftwo
 \else \expandafter \@secondoftwo
 \fi
}%
\providecommand \@ifx [1]{%
 \ifx #1\expandafter \@firstoftwo
 \else \expandafter \@secondoftwo
 \fi
}%
\providecommand \natexlab [1]{#1}%
\providecommand \enquote  [1]{``#1''}%
\providecommand \bibnamefont  [1]{#1}%
\providecommand \bibfnamefont [1]{#1}%
\providecommand \citenamefont [1]{#1}%
\providecommand \href@noop [0]{\@secondoftwo}%
\providecommand \href [0]{\begingroup \@sanitize@url \@href}%
\providecommand \@href[1]{\@@startlink{#1}\@@href}%
\providecommand \@@href[1]{\endgroup#1\@@endlink}%
\providecommand \@sanitize@url [0]{\catcode `\\12\catcode `\$12\catcode
  `\&12\catcode `\#12\catcode `\^12\catcode `\_12\catcode `\%12\relax}%
\providecommand \@@startlink[1]{}%
\providecommand \@@endlink[0]{}%
\providecommand \url  [0]{\begingroup\@sanitize@url \@url }%
\providecommand \@url [1]{\endgroup\@href {#1}{\urlprefix }}%
\providecommand \urlprefix  [0]{URL }%
\providecommand \Eprint [0]{\href }%
\providecommand \doibase [0]{https://doi.org/}%
\providecommand \selectlanguage [0]{\@gobble}%
\providecommand \bibinfo  [0]{\@secondoftwo}%
\providecommand \bibfield  [0]{\@secondoftwo}%
\providecommand \translation [1]{[#1]}%
\providecommand \BibitemOpen [0]{}%
\providecommand \bibitemStop [0]{}%
\providecommand \bibitemNoStop [0]{.\EOS\space}%
\providecommand \EOS [0]{\spacefactor3000\relax}%
\providecommand \BibitemShut  [1]{\csname bibitem#1\endcsname}%
\let\auto@bib@innerbib\@empty
\bibitem [{\citenamefont {Coleman}(2015)}]{coleman2015introduction}%
  \BibitemOpen
  \bibfield  {author} {\bibinfo {author} {\bibfnamefont {P.}~\bibnamefont
  {Coleman}},\ }\href@noop {} {\emph {\bibinfo {title} {Introduction to
  many-body physics}}}\ (\bibinfo  {publisher} {Cambridge University Press},\
  \bibinfo {year} {2015})\BibitemShut {NoStop}%
\bibitem [{\citenamefont {Bernien}\ \emph {et~al.}(2017)\citenamefont
  {Bernien}, \citenamefont {Schwartz}, \citenamefont {Keesling}, \citenamefont
  {Levine}, \citenamefont {Omran}, \citenamefont {Pichler}, \citenamefont
  {Choi}, \citenamefont {Zibrov}, \citenamefont {Endres}, \citenamefont
  {Greiner} \emph {et~al.}}]{bernien2017probing}%
  \BibitemOpen
  \bibfield  {author} {\bibinfo {author} {\bibfnamefont {H.}~\bibnamefont
  {Bernien}}, \bibinfo {author} {\bibfnamefont {S.}~\bibnamefont {Schwartz}},
  \bibinfo {author} {\bibfnamefont {A.}~\bibnamefont {Keesling}}, \bibinfo
  {author} {\bibfnamefont {H.}~\bibnamefont {Levine}}, \bibinfo {author}
  {\bibfnamefont {A.}~\bibnamefont {Omran}}, \bibinfo {author} {\bibfnamefont
  {H.}~\bibnamefont {Pichler}}, \bibinfo {author} {\bibfnamefont
  {S.}~\bibnamefont {Choi}}, \bibinfo {author} {\bibfnamefont {A.~S.}\
  \bibnamefont {Zibrov}}, \bibinfo {author} {\bibfnamefont {M.}~\bibnamefont
  {Endres}}, \bibinfo {author} {\bibfnamefont {M.}~\bibnamefont {Greiner}},
  \emph {et~al.},\ }\bibfield  {title} {\bibinfo {title} {Probing many-body
  dynamics on a 51-atom quantum simulator},\ }\href@noop {} {\bibfield
  {journal} {\bibinfo  {journal} {Nature}\ }\textbf {\bibinfo {volume} {551}},\
  \bibinfo {pages} {579} (\bibinfo {year} {2017})}\BibitemShut {NoStop}%
\bibitem [{\citenamefont {Zhang}\ \emph {et~al.}(2017)\citenamefont {Zhang},
  \citenamefont {Pagano}, \citenamefont {Hess}, \citenamefont {Kyprianidis},
  \citenamefont {Becker}, \citenamefont {Kaplan}, \citenamefont {Gorshkov},
  \citenamefont {Gong},\ and\ \citenamefont {Monroe}}]{zhang2017observation}%
  \BibitemOpen
  \bibfield  {author} {\bibinfo {author} {\bibfnamefont {J.}~\bibnamefont
  {Zhang}}, \bibinfo {author} {\bibfnamefont {G.}~\bibnamefont {Pagano}},
  \bibinfo {author} {\bibfnamefont {P.~W.}\ \bibnamefont {Hess}}, \bibinfo
  {author} {\bibfnamefont {A.}~\bibnamefont {Kyprianidis}}, \bibinfo {author}
  {\bibfnamefont {P.}~\bibnamefont {Becker}}, \bibinfo {author} {\bibfnamefont
  {H.}~\bibnamefont {Kaplan}}, \bibinfo {author} {\bibfnamefont {A.~V.}\
  \bibnamefont {Gorshkov}}, \bibinfo {author} {\bibfnamefont {Z.-X.}\
  \bibnamefont {Gong}},\ and\ \bibinfo {author} {\bibfnamefont
  {C.}~\bibnamefont {Monroe}},\ }\bibfield  {title} {\bibinfo {title}
  {Observation of a many-body dynamical phase transition with a 53-qubit
  quantum simulator},\ }\href@noop {} {\bibfield  {journal} {\bibinfo
  {journal} {Nature}\ }\textbf {\bibinfo {volume} {551}},\ \bibinfo {pages}
  {601} (\bibinfo {year} {2017})}\BibitemShut {NoStop}%
\bibitem [{\citenamefont {Ye}\ \emph {et~al.}(2019)\citenamefont {Ye},
  \citenamefont {Ge}, \citenamefont {Wu}, \citenamefont {Wang}, \citenamefont
  {Gong}, \citenamefont {Zhang}, \citenamefont {Zhu}, \citenamefont {Yang},
  \citenamefont {Li}, \citenamefont {Liang} \emph
  {et~al.}}]{ye2019propagation}%
  \BibitemOpen
  \bibfield  {author} {\bibinfo {author} {\bibfnamefont {Y.}~\bibnamefont
  {Ye}}, \bibinfo {author} {\bibfnamefont {Z.-Y.}\ \bibnamefont {Ge}}, \bibinfo
  {author} {\bibfnamefont {Y.}~\bibnamefont {Wu}}, \bibinfo {author}
  {\bibfnamefont {S.}~\bibnamefont {Wang}}, \bibinfo {author} {\bibfnamefont
  {M.}~\bibnamefont {Gong}}, \bibinfo {author} {\bibfnamefont {Y.-R.}\
  \bibnamefont {Zhang}}, \bibinfo {author} {\bibfnamefont {Q.}~\bibnamefont
  {Zhu}}, \bibinfo {author} {\bibfnamefont {R.}~\bibnamefont {Yang}}, \bibinfo
  {author} {\bibfnamefont {S.}~\bibnamefont {Li}}, \bibinfo {author}
  {\bibfnamefont {F.}~\bibnamefont {Liang}}, \emph {et~al.},\ }\bibfield
  {title} {\bibinfo {title} {Propagation and localization of collective
  excitations on a 24-qubit superconducting processor},\ }\href@noop {}
  {\bibfield  {journal} {\bibinfo  {journal} {Physical Review Letters}\
  }\textbf {\bibinfo {volume} {123}},\ \bibinfo {pages} {050502} (\bibinfo
  {year} {2019})}\BibitemShut {NoStop}%
\bibitem [{\citenamefont {Chen}\ \emph {et~al.}(2021)\citenamefont {Chen},
  \citenamefont {Sun}, \citenamefont {Gong}, \citenamefont {Zhu}, \citenamefont
  {Zhang}, \citenamefont {Wu}, \citenamefont {Ye}, \citenamefont {Zha},
  \citenamefont {Li}, \citenamefont {Guo} \emph
  {et~al.}}]{chen2021observation}%
  \BibitemOpen
  \bibfield  {author} {\bibinfo {author} {\bibfnamefont {F.}~\bibnamefont
  {Chen}}, \bibinfo {author} {\bibfnamefont {Z.-H.}\ \bibnamefont {Sun}},
  \bibinfo {author} {\bibfnamefont {M.}~\bibnamefont {Gong}}, \bibinfo {author}
  {\bibfnamefont {Q.}~\bibnamefont {Zhu}}, \bibinfo {author} {\bibfnamefont
  {Y.-R.}\ \bibnamefont {Zhang}}, \bibinfo {author} {\bibfnamefont
  {Y.}~\bibnamefont {Wu}}, \bibinfo {author} {\bibfnamefont {Y.}~\bibnamefont
  {Ye}}, \bibinfo {author} {\bibfnamefont {C.}~\bibnamefont {Zha}}, \bibinfo
  {author} {\bibfnamefont {S.}~\bibnamefont {Li}}, \bibinfo {author}
  {\bibfnamefont {S.}~\bibnamefont {Guo}}, \emph {et~al.},\ }\bibfield  {title}
  {\bibinfo {title} {Observation of strong and weak thermalization in a
  superconducting quantum processor},\ }\href@noop {} {\bibfield  {journal}
  {\bibinfo  {journal} {Physical Review Letters}\ }\textbf {\bibinfo {volume}
  {127}},\ \bibinfo {pages} {020602} (\bibinfo {year} {2021})}\BibitemShut
  {NoStop}%
\bibitem [{\citenamefont {Gong}\ \emph
  {et~al.}(2021{\natexlab{a}})\citenamefont {Gong}, \citenamefont
  {de~Moraes~Neto}, \citenamefont {Zha}, \citenamefont {Wu}, \citenamefont
  {Rong}, \citenamefont {Ye}, \citenamefont {Li}, \citenamefont {Zhu},
  \citenamefont {Wang}, \citenamefont {Zhao} \emph
  {et~al.}}]{gong2021experimental}%
  \BibitemOpen
  \bibfield  {author} {\bibinfo {author} {\bibfnamefont {M.}~\bibnamefont
  {Gong}}, \bibinfo {author} {\bibfnamefont {G.~D.}\ \bibnamefont
  {de~Moraes~Neto}}, \bibinfo {author} {\bibfnamefont {C.}~\bibnamefont {Zha}},
  \bibinfo {author} {\bibfnamefont {Y.}~\bibnamefont {Wu}}, \bibinfo {author}
  {\bibfnamefont {H.}~\bibnamefont {Rong}}, \bibinfo {author} {\bibfnamefont
  {Y.}~\bibnamefont {Ye}}, \bibinfo {author} {\bibfnamefont {S.}~\bibnamefont
  {Li}}, \bibinfo {author} {\bibfnamefont {Q.}~\bibnamefont {Zhu}}, \bibinfo
  {author} {\bibfnamefont {S.}~\bibnamefont {Wang}}, \bibinfo {author}
  {\bibfnamefont {Y.}~\bibnamefont {Zhao}}, \emph {et~al.},\ }\bibfield
  {title} {\bibinfo {title} {Experimental characterization of the quantum
  many-body localization transition},\ }\href@noop {} {\bibfield  {journal}
  {\bibinfo  {journal} {Physical Review Research}\ }\textbf {\bibinfo {volume}
  {3}},\ \bibinfo {pages} {033043} (\bibinfo {year}
  {2021}{\natexlab{a}})}\BibitemShut {NoStop}%
\bibitem [{\citenamefont {Ebadi}\ \emph {et~al.}(2021)\citenamefont {Ebadi},
  \citenamefont {Wang}, \citenamefont {Levine}, \citenamefont {Keesling},
  \citenamefont {Semeghini}, \citenamefont {Omran}, \citenamefont {Bluvstein},
  \citenamefont {Samajdar}, \citenamefont {Pichler}, \citenamefont {Ho} \emph
  {et~al.}}]{ebadi2021quantum}%
  \BibitemOpen
  \bibfield  {author} {\bibinfo {author} {\bibfnamefont {S.}~\bibnamefont
  {Ebadi}}, \bibinfo {author} {\bibfnamefont {T.~T.}\ \bibnamefont {Wang}},
  \bibinfo {author} {\bibfnamefont {H.}~\bibnamefont {Levine}}, \bibinfo
  {author} {\bibfnamefont {A.}~\bibnamefont {Keesling}}, \bibinfo {author}
  {\bibfnamefont {G.}~\bibnamefont {Semeghini}}, \bibinfo {author}
  {\bibfnamefont {A.}~\bibnamefont {Omran}}, \bibinfo {author} {\bibfnamefont
  {D.}~\bibnamefont {Bluvstein}}, \bibinfo {author} {\bibfnamefont
  {R.}~\bibnamefont {Samajdar}}, \bibinfo {author} {\bibfnamefont
  {H.}~\bibnamefont {Pichler}}, \bibinfo {author} {\bibfnamefont {W.~W.}\
  \bibnamefont {Ho}}, \emph {et~al.},\ }\bibfield  {title} {\bibinfo {title}
  {Quantum phases of matter on a 256-atom programmable quantum simulator},\
  }\href@noop {} {\bibfield  {journal} {\bibinfo  {journal} {Nature}\ }\textbf
  {\bibinfo {volume} {595}},\ \bibinfo {pages} {227} (\bibinfo {year}
  {2021})}\BibitemShut {NoStop}%
\bibitem [{\citenamefont {Eisert}\ \emph {et~al.}(2015)\citenamefont {Eisert},
  \citenamefont {Friesdorf},\ and\ \citenamefont
  {Gogolin}}]{eisert2015quantum}%
  \BibitemOpen
  \bibfield  {author} {\bibinfo {author} {\bibfnamefont {J.}~\bibnamefont
  {Eisert}}, \bibinfo {author} {\bibfnamefont {M.}~\bibnamefont {Friesdorf}},\
  and\ \bibinfo {author} {\bibfnamefont {C.}~\bibnamefont {Gogolin}},\
  }\bibfield  {title} {\bibinfo {title} {Quantum many-body systems out of
  equilibrium},\ }\href@noop {} {\bibfield  {journal} {\bibinfo  {journal}
  {Nature Physics}\ }\textbf {\bibinfo {volume} {11}},\ \bibinfo {pages} {124}
  (\bibinfo {year} {2015})}\BibitemShut {NoStop}%
\bibitem [{\citenamefont {Schweigler}\ \emph {et~al.}(2017)\citenamefont
  {Schweigler}, \citenamefont {Kasper}, \citenamefont {Erne}, \citenamefont
  {Mazets}, \citenamefont {Rauer}, \citenamefont {Cataldini}, \citenamefont
  {Langen}, \citenamefont {Gasenzer}, \citenamefont {Berges},\ and\
  \citenamefont {Schmiedmayer}}]{schweigler2017experimental}%
  \BibitemOpen
  \bibfield  {author} {\bibinfo {author} {\bibfnamefont {T.}~\bibnamefont
  {Schweigler}}, \bibinfo {author} {\bibfnamefont {V.}~\bibnamefont {Kasper}},
  \bibinfo {author} {\bibfnamefont {S.}~\bibnamefont {Erne}}, \bibinfo {author}
  {\bibfnamefont {I.}~\bibnamefont {Mazets}}, \bibinfo {author} {\bibfnamefont
  {B.}~\bibnamefont {Rauer}}, \bibinfo {author} {\bibfnamefont
  {F.}~\bibnamefont {Cataldini}}, \bibinfo {author} {\bibfnamefont
  {T.}~\bibnamefont {Langen}}, \bibinfo {author} {\bibfnamefont
  {T.}~\bibnamefont {Gasenzer}}, \bibinfo {author} {\bibfnamefont
  {J.}~\bibnamefont {Berges}},\ and\ \bibinfo {author} {\bibfnamefont
  {J.}~\bibnamefont {Schmiedmayer}},\ }\bibfield  {title} {\bibinfo {title}
  {Experimental characterization of a quantum many-body system via higher-order
  correlations},\ }\href@noop {} {\bibfield  {journal} {\bibinfo  {journal}
  {Nature}\ }\textbf {\bibinfo {volume} {545}},\ \bibinfo {pages} {323}
  (\bibinfo {year} {2017})}\BibitemShut {NoStop}%
\bibitem [{\citenamefont {Lanyon}\ \emph {et~al.}(2017)\citenamefont {Lanyon},
  \citenamefont {Maier}, \citenamefont {Holz{\"a}pfel}, \citenamefont
  {Baumgratz}, \citenamefont {Hempel}, \citenamefont {Jurcevic}, \citenamefont
  {Dhand}, \citenamefont {Buyskikh}, \citenamefont {Daley}, \citenamefont
  {Cramer} \emph {et~al.}}]{lanyon2017efficient}%
  \BibitemOpen
  \bibfield  {author} {\bibinfo {author} {\bibfnamefont {B.}~\bibnamefont
  {Lanyon}}, \bibinfo {author} {\bibfnamefont {C.}~\bibnamefont {Maier}},
  \bibinfo {author} {\bibfnamefont {M.}~\bibnamefont {Holz{\"a}pfel}}, \bibinfo
  {author} {\bibfnamefont {T.}~\bibnamefont {Baumgratz}}, \bibinfo {author}
  {\bibfnamefont {C.}~\bibnamefont {Hempel}}, \bibinfo {author} {\bibfnamefont
  {P.}~\bibnamefont {Jurcevic}}, \bibinfo {author} {\bibfnamefont
  {I.}~\bibnamefont {Dhand}}, \bibinfo {author} {\bibfnamefont
  {A.}~\bibnamefont {Buyskikh}}, \bibinfo {author} {\bibfnamefont
  {A.}~\bibnamefont {Daley}}, \bibinfo {author} {\bibfnamefont
  {M.}~\bibnamefont {Cramer}}, \emph {et~al.},\ }\bibfield  {title} {\bibinfo
  {title} {Efficient tomography of a quantum many-body system},\ }\href@noop {}
  {\bibfield  {journal} {\bibinfo  {journal} {Nature Physics}\ }\textbf
  {\bibinfo {volume} {13}},\ \bibinfo {pages} {1158} (\bibinfo {year}
  {2017})}\BibitemShut {NoStop}%
\bibitem [{\citenamefont {Zhao}\ \emph {et~al.}(2020)\citenamefont {Zhao},
  \citenamefont {Vovrosh}, \citenamefont {Mintert},\ and\ \citenamefont
  {Knolle}}]{zhao2020quantum}%
  \BibitemOpen
  \bibfield  {author} {\bibinfo {author} {\bibfnamefont {H.}~\bibnamefont
  {Zhao}}, \bibinfo {author} {\bibfnamefont {J.}~\bibnamefont {Vovrosh}},
  \bibinfo {author} {\bibfnamefont {F.}~\bibnamefont {Mintert}},\ and\ \bibinfo
  {author} {\bibfnamefont {J.}~\bibnamefont {Knolle}},\ }\bibfield  {title}
  {\bibinfo {title} {Quantum many-body scars in optical lattices},\ }\href@noop
  {} {\bibfield  {journal} {\bibinfo  {journal} {Physical review letters}\
  }\textbf {\bibinfo {volume} {124}},\ \bibinfo {pages} {160604} (\bibinfo
  {year} {2020})}\BibitemShut {NoStop}%
\bibitem [{\citenamefont {Thomas}\ \emph {et~al.}(2018)\citenamefont {Thomas},
  \citenamefont {Lippe}, \citenamefont {Eichert},\ and\ \citenamefont
  {Ott}}]{thomas2018experimental}%
  \BibitemOpen
  \bibfield  {author} {\bibinfo {author} {\bibfnamefont {O.}~\bibnamefont
  {Thomas}}, \bibinfo {author} {\bibfnamefont {C.}~\bibnamefont {Lippe}},
  \bibinfo {author} {\bibfnamefont {T.}~\bibnamefont {Eichert}},\ and\ \bibinfo
  {author} {\bibfnamefont {H.}~\bibnamefont {Ott}},\ }\bibfield  {title}
  {\bibinfo {title} {Experimental realization of a rydberg optical feshbach
  resonance in a quantum many-body system},\ }\href@noop {} {\bibfield
  {journal} {\bibinfo  {journal} {Nature communications}\ }\textbf {\bibinfo
  {volume} {9}},\ \bibinfo {pages} {1} (\bibinfo {year} {2018})}\BibitemShut
  {NoStop}%
\bibitem [{\citenamefont {Pr{\"u}fer}\ \emph {et~al.}(2020)\citenamefont
  {Pr{\"u}fer}, \citenamefont {Zache}, \citenamefont {Kunkel}, \citenamefont
  {Lannig}, \citenamefont {Bonnin}, \citenamefont {Strobel}, \citenamefont
  {Berges},\ and\ \citenamefont {Oberthaler}}]{prufer2020experimental}%
  \BibitemOpen
  \bibfield  {author} {\bibinfo {author} {\bibfnamefont {M.}~\bibnamefont
  {Pr{\"u}fer}}, \bibinfo {author} {\bibfnamefont {T.~V.}\ \bibnamefont
  {Zache}}, \bibinfo {author} {\bibfnamefont {P.}~\bibnamefont {Kunkel}},
  \bibinfo {author} {\bibfnamefont {S.}~\bibnamefont {Lannig}}, \bibinfo
  {author} {\bibfnamefont {A.}~\bibnamefont {Bonnin}}, \bibinfo {author}
  {\bibfnamefont {H.}~\bibnamefont {Strobel}}, \bibinfo {author} {\bibfnamefont
  {J.}~\bibnamefont {Berges}},\ and\ \bibinfo {author} {\bibfnamefont {M.~K.}\
  \bibnamefont {Oberthaler}},\ }\bibfield  {title} {\bibinfo {title}
  {Experimental extraction of the quantum effective action for a
  non-equilibrium many-body system},\ }\href@noop {} {\bibfield  {journal}
  {\bibinfo  {journal} {Nature Physics}\ }\textbf {\bibinfo {volume} {16}},\
  \bibinfo {pages} {1012} (\bibinfo {year} {2020})}\BibitemShut {NoStop}%
\bibitem [{\citenamefont {Guo}\ \emph {et~al.}(2021)\citenamefont {Guo},
  \citenamefont {Cheng}, \citenamefont {Sun}, \citenamefont {Song},
  \citenamefont {Li}, \citenamefont {Wang}, \citenamefont {Ren}, \citenamefont
  {Dong}, \citenamefont {Zheng}, \citenamefont {Zhang} \emph
  {et~al.}}]{guo2021observation}%
  \BibitemOpen
  \bibfield  {author} {\bibinfo {author} {\bibfnamefont {Q.}~\bibnamefont
  {Guo}}, \bibinfo {author} {\bibfnamefont {C.}~\bibnamefont {Cheng}}, \bibinfo
  {author} {\bibfnamefont {Z.-H.}\ \bibnamefont {Sun}}, \bibinfo {author}
  {\bibfnamefont {Z.}~\bibnamefont {Song}}, \bibinfo {author} {\bibfnamefont
  {H.}~\bibnamefont {Li}}, \bibinfo {author} {\bibfnamefont {Z.}~\bibnamefont
  {Wang}}, \bibinfo {author} {\bibfnamefont {W.}~\bibnamefont {Ren}}, \bibinfo
  {author} {\bibfnamefont {H.}~\bibnamefont {Dong}}, \bibinfo {author}
  {\bibfnamefont {D.}~\bibnamefont {Zheng}}, \bibinfo {author} {\bibfnamefont
  {Y.-R.}\ \bibnamefont {Zhang}}, \emph {et~al.},\ }\bibfield  {title}
  {\bibinfo {title} {Observation of energy-resolved many-body localization},\
  }\href@noop {} {\bibfield  {journal} {\bibinfo  {journal} {Nature Physics}\
  }\textbf {\bibinfo {volume} {17}},\ \bibinfo {pages} {234} (\bibinfo {year}
  {2021})}\BibitemShut {NoStop}%
\bibitem [{\citenamefont {Savary}\ and\ \citenamefont
  {Balents}(2016)}]{savary2016quantum}%
  \BibitemOpen
  \bibfield  {author} {\bibinfo {author} {\bibfnamefont {L.}~\bibnamefont
  {Savary}}\ and\ \bibinfo {author} {\bibfnamefont {L.}~\bibnamefont
  {Balents}},\ }\bibfield  {title} {\bibinfo {title} {Quantum spin liquids: a
  review},\ }\href@noop {} {\bibfield  {journal} {\bibinfo  {journal} {Reports
  on Progress in Physics}\ }\textbf {\bibinfo {volume} {80}},\ \bibinfo {pages}
  {016502} (\bibinfo {year} {2016})}\BibitemShut {NoStop}%
\bibitem [{\citenamefont {Wen}(1992)}]{wen1992theory}%
  \BibitemOpen
  \bibfield  {author} {\bibinfo {author} {\bibfnamefont {X.-G.}\ \bibnamefont
  {Wen}},\ }\bibfield  {title} {\bibinfo {title} {Theory of the edge states in
  fractional quantum hall effects},\ }\href@noop {} {\bibfield  {journal}
  {\bibinfo  {journal} {International Journal of Modern Physics B}\ }\textbf
  {\bibinfo {volume} {6}},\ \bibinfo {pages} {1711} (\bibinfo {year}
  {1992})}\BibitemShut {NoStop}%
\bibitem [{\citenamefont {Felser}\ \emph {et~al.}(2021)\citenamefont {Felser},
  \citenamefont {Notarnicola},\ and\ \citenamefont
  {Montangero}}]{felser2021efficient}%
  \BibitemOpen
  \bibfield  {author} {\bibinfo {author} {\bibfnamefont {T.}~\bibnamefont
  {Felser}}, \bibinfo {author} {\bibfnamefont {S.}~\bibnamefont
  {Notarnicola}},\ and\ \bibinfo {author} {\bibfnamefont {S.}~\bibnamefont
  {Montangero}},\ }\bibfield  {title} {\bibinfo {title} {Efficient tensor
  network ansatz for high-dimensional quantum many-body problems},\ }\href@noop
  {} {\bibfield  {journal} {\bibinfo  {journal} {Physical Review Letters}\
  }\textbf {\bibinfo {volume} {126}},\ \bibinfo {pages} {170603} (\bibinfo
  {year} {2021})}\BibitemShut {NoStop}%
\bibitem [{\citenamefont {Melko}\ \emph {et~al.}(2019)\citenamefont {Melko},
  \citenamefont {Carleo}, \citenamefont {Carrasquilla},\ and\ \citenamefont
  {Cirac}}]{melko2019restricted}%
  \BibitemOpen
  \bibfield  {author} {\bibinfo {author} {\bibfnamefont {R.~G.}\ \bibnamefont
  {Melko}}, \bibinfo {author} {\bibfnamefont {G.}~\bibnamefont {Carleo}},
  \bibinfo {author} {\bibfnamefont {J.}~\bibnamefont {Carrasquilla}},\ and\
  \bibinfo {author} {\bibfnamefont {J.~I.}\ \bibnamefont {Cirac}},\ }\bibfield
  {title} {\bibinfo {title} {Restricted boltzmann machines in quantum
  physics},\ }\href@noop {} {\bibfield  {journal} {\bibinfo  {journal} {Nature
  Physics}\ }\textbf {\bibinfo {volume} {15}},\ \bibinfo {pages} {887}
  (\bibinfo {year} {2019})}\BibitemShut {NoStop}%
\bibitem [{\citenamefont {Carleo}\ \emph {et~al.}(2018)\citenamefont {Carleo},
  \citenamefont {Nomura},\ and\ \citenamefont
  {Imada}}]{carleo2018constructing}%
  \BibitemOpen
  \bibfield  {author} {\bibinfo {author} {\bibfnamefont {G.}~\bibnamefont
  {Carleo}}, \bibinfo {author} {\bibfnamefont {Y.}~\bibnamefont {Nomura}},\
  and\ \bibinfo {author} {\bibfnamefont {M.}~\bibnamefont {Imada}},\ }\bibfield
   {title} {\bibinfo {title} {Constructing exact representations of quantum
  many-body systems with deep neural networks},\ }\href@noop {} {\bibfield
  {journal} {\bibinfo  {journal} {Nature Communications}\ }\textbf {\bibinfo
  {volume} {9}},\ \bibinfo {pages} {1} (\bibinfo {year} {2018})}\BibitemShut
  {NoStop}%
\bibitem [{\citenamefont {Vicentini}(2021)}]{vicentini2021machine}%
  \BibitemOpen
  \bibfield  {author} {\bibinfo {author} {\bibfnamefont {F.}~\bibnamefont
  {Vicentini}},\ }\bibfield  {title} {\bibinfo {title} {Machine learning
  toolbox for quantum many body physics},\ }\href@noop {} {\bibfield  {journal}
  {\bibinfo  {journal} {Nature Reviews Physics}\ }\textbf {\bibinfo {volume}
  {3}},\ \bibinfo {pages} {156} (\bibinfo {year} {2021})}\BibitemShut {NoStop}%
\bibitem [{\citenamefont {Gong}\ \emph
  {et~al.}(2021{\natexlab{b}})\citenamefont {Gong}, \citenamefont {Wang},
  \citenamefont {Zha}, \citenamefont {Chen}, \citenamefont {Huang},
  \citenamefont {Wu}, \citenamefont {Zhu}, \citenamefont {Zhao}, \citenamefont
  {Li}, \citenamefont {Guo} \emph {et~al.}}]{gong2021quantum}%
  \BibitemOpen
  \bibfield  {author} {\bibinfo {author} {\bibfnamefont {M.}~\bibnamefont
  {Gong}}, \bibinfo {author} {\bibfnamefont {S.}~\bibnamefont {Wang}}, \bibinfo
  {author} {\bibfnamefont {C.}~\bibnamefont {Zha}}, \bibinfo {author}
  {\bibfnamefont {M.-C.}\ \bibnamefont {Chen}}, \bibinfo {author}
  {\bibfnamefont {H.-L.}\ \bibnamefont {Huang}}, \bibinfo {author}
  {\bibfnamefont {Y.}~\bibnamefont {Wu}}, \bibinfo {author} {\bibfnamefont
  {Q.}~\bibnamefont {Zhu}}, \bibinfo {author} {\bibfnamefont {Y.}~\bibnamefont
  {Zhao}}, \bibinfo {author} {\bibfnamefont {S.}~\bibnamefont {Li}}, \bibinfo
  {author} {\bibfnamefont {S.}~\bibnamefont {Guo}}, \emph {et~al.},\ }\bibfield
   {title} {\bibinfo {title} {Quantum walks on a programmable two-dimensional
  62-qubit superconducting processor},\ }\href@noop {} {\bibfield  {journal}
  {\bibinfo  {journal} {Science}\ }\textbf {\bibinfo {volume} {372}},\ \bibinfo
  {pages} {948} (\bibinfo {year} {2021}{\natexlab{b}})}\BibitemShut {NoStop}%
\bibitem [{\citenamefont {Wu}\ \emph {et~al.}(2021)\citenamefont {Wu},
  \citenamefont {Bao}, \citenamefont {Cao}, \citenamefont {Chen}, \citenamefont
  {Chen}, \citenamefont {Chen}, \citenamefont {Chung}, \citenamefont {Deng},
  \citenamefont {Du}, \citenamefont {Fan}, \citenamefont {Gong}, \citenamefont
  {Guo}, \citenamefont {Guo}, \citenamefont {Guo}, \citenamefont {Han},
  \citenamefont {Hong}, \citenamefont {Huang}, \citenamefont {Huo},
  \citenamefont {Li}, \citenamefont {Li}, \citenamefont {Li}, \citenamefont
  {Li}, \citenamefont {Liang}, \citenamefont {Lin}, \citenamefont {Lin},
  \citenamefont {Qian}, \citenamefont {Qiao}, \citenamefont {Rong},
  \citenamefont {Su}, \citenamefont {Sun}, \citenamefont {Wang}, \citenamefont
  {Wang}, \citenamefont {Wu}, \citenamefont {Xu}, \citenamefont {Yan},
  \citenamefont {Yang}, \citenamefont {Yang}, \citenamefont {Ye}, \citenamefont
  {Yin}, \citenamefont {Ying}, \citenamefont {Yu}, \citenamefont {Zha},
  \citenamefont {Zhang}, \citenamefont {Zhang}, \citenamefont {Zhang},
  \citenamefont {Zhang}, \citenamefont {Zhao}, \citenamefont {Zhao},
  \citenamefont {Zhou}, \citenamefont {Zhu}, \citenamefont {Lu}, \citenamefont
  {Peng}, \citenamefont {Zhu},\ and\ \citenamefont {Pan}}]{wu2021strong}%
  \BibitemOpen
  \bibfield  {author} {\bibinfo {author} {\bibfnamefont {Y.}~\bibnamefont
  {Wu}}, \bibinfo {author} {\bibfnamefont {W.-S.}\ \bibnamefont {Bao}},
  \bibinfo {author} {\bibfnamefont {S.}~\bibnamefont {Cao}}, \bibinfo {author}
  {\bibfnamefont {F.}~\bibnamefont {Chen}}, \bibinfo {author} {\bibfnamefont
  {M.-C.}\ \bibnamefont {Chen}}, \bibinfo {author} {\bibfnamefont
  {X.}~\bibnamefont {Chen}}, \bibinfo {author} {\bibfnamefont {T.-H.}\
  \bibnamefont {Chung}}, \bibinfo {author} {\bibfnamefont {H.}~\bibnamefont
  {Deng}}, \bibinfo {author} {\bibfnamefont {Y.}~\bibnamefont {Du}}, \bibinfo
  {author} {\bibfnamefont {D.}~\bibnamefont {Fan}}, \bibinfo {author}
  {\bibfnamefont {M.}~\bibnamefont {Gong}}, \bibinfo {author} {\bibfnamefont
  {C.}~\bibnamefont {Guo}}, \bibinfo {author} {\bibfnamefont {C.}~\bibnamefont
  {Guo}}, \bibinfo {author} {\bibfnamefont {S.}~\bibnamefont {Guo}}, \bibinfo
  {author} {\bibfnamefont {L.}~\bibnamefont {Han}}, \bibinfo {author}
  {\bibfnamefont {L.}~\bibnamefont {Hong}}, \bibinfo {author} {\bibfnamefont
  {H.-L.}\ \bibnamefont {Huang}}, \bibinfo {author} {\bibfnamefont {Y.-H.}\
  \bibnamefont {Huo}}, \bibinfo {author} {\bibfnamefont {L.}~\bibnamefont
  {Li}}, \bibinfo {author} {\bibfnamefont {N.}~\bibnamefont {Li}}, \bibinfo
  {author} {\bibfnamefont {S.}~\bibnamefont {Li}}, \bibinfo {author}
  {\bibfnamefont {Y.}~\bibnamefont {Li}}, \bibinfo {author} {\bibfnamefont
  {F.}~\bibnamefont {Liang}}, \bibinfo {author} {\bibfnamefont
  {C.}~\bibnamefont {Lin}}, \bibinfo {author} {\bibfnamefont {J.}~\bibnamefont
  {Lin}}, \bibinfo {author} {\bibfnamefont {H.}~\bibnamefont {Qian}}, \bibinfo
  {author} {\bibfnamefont {D.}~\bibnamefont {Qiao}}, \bibinfo {author}
  {\bibfnamefont {H.}~\bibnamefont {Rong}}, \bibinfo {author} {\bibfnamefont
  {H.}~\bibnamefont {Su}}, \bibinfo {author} {\bibfnamefont {L.}~\bibnamefont
  {Sun}}, \bibinfo {author} {\bibfnamefont {L.}~\bibnamefont {Wang}}, \bibinfo
  {author} {\bibfnamefont {S.}~\bibnamefont {Wang}}, \bibinfo {author}
  {\bibfnamefont {D.}~\bibnamefont {Wu}}, \bibinfo {author} {\bibfnamefont
  {Y.}~\bibnamefont {Xu}}, \bibinfo {author} {\bibfnamefont {K.}~\bibnamefont
  {Yan}}, \bibinfo {author} {\bibfnamefont {W.}~\bibnamefont {Yang}}, \bibinfo
  {author} {\bibfnamefont {Y.}~\bibnamefont {Yang}}, \bibinfo {author}
  {\bibfnamefont {Y.}~\bibnamefont {Ye}}, \bibinfo {author} {\bibfnamefont
  {J.}~\bibnamefont {Yin}}, \bibinfo {author} {\bibfnamefont {C.}~\bibnamefont
  {Ying}}, \bibinfo {author} {\bibfnamefont {J.}~\bibnamefont {Yu}}, \bibinfo
  {author} {\bibfnamefont {C.}~\bibnamefont {Zha}}, \bibinfo {author}
  {\bibfnamefont {C.}~\bibnamefont {Zhang}}, \bibinfo {author} {\bibfnamefont
  {H.}~\bibnamefont {Zhang}}, \bibinfo {author} {\bibfnamefont
  {K.}~\bibnamefont {Zhang}}, \bibinfo {author} {\bibfnamefont
  {Y.}~\bibnamefont {Zhang}}, \bibinfo {author} {\bibfnamefont
  {H.}~\bibnamefont {Zhao}}, \bibinfo {author} {\bibfnamefont {Y.}~\bibnamefont
  {Zhao}}, \bibinfo {author} {\bibfnamefont {L.}~\bibnamefont {Zhou}}, \bibinfo
  {author} {\bibfnamefont {Q.}~\bibnamefont {Zhu}}, \bibinfo {author}
  {\bibfnamefont {C.-Y.}\ \bibnamefont {Lu}}, \bibinfo {author} {\bibfnamefont
  {C.-Z.}\ \bibnamefont {Peng}}, \bibinfo {author} {\bibfnamefont
  {X.}~\bibnamefont {Zhu}},\ and\ \bibinfo {author} {\bibfnamefont {J.-W.}\
  \bibnamefont {Pan}},\ }\bibfield  {title} {\bibinfo {title} {{Strong quantum
  computational advantage using a superconducting quantum processor}},\ }\href
  {https://doi.org/10.1103/PhysRevLett.127.180501} {\bibfield  {journal}
  {\bibinfo  {journal} {Physical Review Letters}\ }\textbf {\bibinfo {volume}
  {127}},\ \bibinfo {pages} {180501} (\bibinfo {year} {2021})}\BibitemShut
  {NoStop}%
\bibitem [{\citenamefont {Zhu}\ \emph {et~al.}(2021)\citenamefont {Zhu},
  \citenamefont {Cao}, \citenamefont {Chen}, \citenamefont {Chen},
  \citenamefont {Chen}, \citenamefont {Chung}, \citenamefont {Deng},
  \citenamefont {Du}, \citenamefont {Fan}, \citenamefont {Gong} \emph
  {et~al.}}]{zhu2021quantum}%
  \BibitemOpen
  \bibfield  {author} {\bibinfo {author} {\bibfnamefont {Q.}~\bibnamefont
  {Zhu}}, \bibinfo {author} {\bibfnamefont {S.}~\bibnamefont {Cao}}, \bibinfo
  {author} {\bibfnamefont {F.}~\bibnamefont {Chen}}, \bibinfo {author}
  {\bibfnamefont {M.-C.}\ \bibnamefont {Chen}}, \bibinfo {author}
  {\bibfnamefont {X.}~\bibnamefont {Chen}}, \bibinfo {author} {\bibfnamefont
  {T.-H.}\ \bibnamefont {Chung}}, \bibinfo {author} {\bibfnamefont
  {H.}~\bibnamefont {Deng}}, \bibinfo {author} {\bibfnamefont {Y.}~\bibnamefont
  {Du}}, \bibinfo {author} {\bibfnamefont {D.}~\bibnamefont {Fan}}, \bibinfo
  {author} {\bibfnamefont {M.}~\bibnamefont {Gong}}, \emph {et~al.},\
  }\bibfield  {title} {\bibinfo {title} {Quantum computational advantage via
  60-qubit 24-cycle random circuit sampling},\ }\href@noop {} {\bibfield
  {journal} {\bibinfo  {journal} {arXiv:2109.03494}\ } (\bibinfo {year}
  {2021})}\BibitemShut {NoStop}%
\bibitem [{\citenamefont {Gao}\ and\ \citenamefont
  {Duan}(2017)}]{gao2017efficient}%
  \BibitemOpen
  \bibfield  {author} {\bibinfo {author} {\bibfnamefont {X.}~\bibnamefont
  {Gao}}\ and\ \bibinfo {author} {\bibfnamefont {L.-M.}\ \bibnamefont {Duan}},\
  }\bibfield  {title} {\bibinfo {title} {Efficient representation of quantum
  many-body states with deep neural networks},\ }\href@noop {} {\bibfield
  {journal} {\bibinfo  {journal} {Nature Communications}\ }\textbf {\bibinfo
  {volume} {8}},\ \bibinfo {pages} {1} (\bibinfo {year} {2017})}\BibitemShut
  {NoStop}%
\bibitem [{\citenamefont {Cai}\ and\ \citenamefont
  {Liu}(2018)}]{cai2018approximating}%
  \BibitemOpen
  \bibfield  {author} {\bibinfo {author} {\bibfnamefont {Z.}~\bibnamefont
  {Cai}}\ and\ \bibinfo {author} {\bibfnamefont {J.}~\bibnamefont {Liu}},\
  }\bibfield  {title} {\bibinfo {title} {Approximating quantum many-body wave
  functions using artificial neural networks},\ }\href@noop {} {\bibfield
  {journal} {\bibinfo  {journal} {Physical Review B}\ }\textbf {\bibinfo
  {volume} {97}},\ \bibinfo {pages} {035116} (\bibinfo {year}
  {2018})}\BibitemShut {NoStop}%
\bibitem [{\citenamefont {Bravyi}\ \emph {et~al.}(2019)\citenamefont {Bravyi},
  \citenamefont {Gosset}, \citenamefont {K{\"o}nig},\ and\ \citenamefont
  {Temme}}]{bravyi2019approximation}%
  \BibitemOpen
  \bibfield  {author} {\bibinfo {author} {\bibfnamefont {S.}~\bibnamefont
  {Bravyi}}, \bibinfo {author} {\bibfnamefont {D.}~\bibnamefont {Gosset}},
  \bibinfo {author} {\bibfnamefont {R.}~\bibnamefont {K{\"o}nig}},\ and\
  \bibinfo {author} {\bibfnamefont {K.}~\bibnamefont {Temme}},\ }\bibfield
  {title} {\bibinfo {title} {Approximation algorithms for quantum many-body
  problems},\ }\href@noop {} {\bibfield  {journal} {\bibinfo  {journal}
  {Journal of Mathematical Physics}\ }\textbf {\bibinfo {volume} {60}},\
  \bibinfo {pages} {032203} (\bibinfo {year} {2019})}\BibitemShut {NoStop}%
\bibitem [{\citenamefont {Carleo}\ and\ \citenamefont
  {Troyer}(2017)}]{carleo2017solving}%
  \BibitemOpen
  \bibfield  {author} {\bibinfo {author} {\bibfnamefont {G.}~\bibnamefont
  {Carleo}}\ and\ \bibinfo {author} {\bibfnamefont {M.}~\bibnamefont
  {Troyer}},\ }\bibfield  {title} {\bibinfo {title} {Solving the quantum
  many-body problem with artificial neural networks},\ }\href@noop {}
  {\bibfield  {journal} {\bibinfo  {journal} {Science}\ }\textbf {\bibinfo
  {volume} {355}},\ \bibinfo {pages} {602} (\bibinfo {year}
  {2017})}\BibitemShut {NoStop}%
\bibitem [{\citenamefont {Carrasquilla}\ and\ \citenamefont
  {Melko}(2017)}]{carrasquilla2017machine}%
  \BibitemOpen
  \bibfield  {author} {\bibinfo {author} {\bibfnamefont {J.}~\bibnamefont
  {Carrasquilla}}\ and\ \bibinfo {author} {\bibfnamefont {R.~G.}\ \bibnamefont
  {Melko}},\ }\bibfield  {title} {\bibinfo {title} {Machine learning phases of
  matter},\ }\href@noop {} {\bibfield  {journal} {\bibinfo  {journal} {Nature
  Physics}\ }\textbf {\bibinfo {volume} {13}},\ \bibinfo {pages} {431}
  (\bibinfo {year} {2017})}\BibitemShut {NoStop}%
\bibitem [{\citenamefont {Vidal}(2004)}]{vidal2004efficient}%
  \BibitemOpen
  \bibfield  {author} {\bibinfo {author} {\bibfnamefont {G.}~\bibnamefont
  {Vidal}},\ }\bibfield  {title} {\bibinfo {title} {Efficient simulation of
  one-dimensional quantum many-body systems},\ }\href@noop {} {\bibfield
  {journal} {\bibinfo  {journal} {Physical Review Letters}\ }\textbf {\bibinfo
  {volume} {93}},\ \bibinfo {pages} {040502} (\bibinfo {year}
  {2004})}\BibitemShut {NoStop}%
\bibitem [{\citenamefont {Cramer}\ \emph {et~al.}(2010)\citenamefont {Cramer},
  \citenamefont {Plenio}, \citenamefont {Flammia}, \citenamefont {Somma},
  \citenamefont {Gross}, \citenamefont {Bartlett}, \citenamefont
  {Landon-Cardinal}, \citenamefont {Poulin},\ and\ \citenamefont
  {Liu}}]{cramer2010efficient}%
  \BibitemOpen
  \bibfield  {author} {\bibinfo {author} {\bibfnamefont {M.}~\bibnamefont
  {Cramer}}, \bibinfo {author} {\bibfnamefont {M.~B.}\ \bibnamefont {Plenio}},
  \bibinfo {author} {\bibfnamefont {S.~T.}\ \bibnamefont {Flammia}}, \bibinfo
  {author} {\bibfnamefont {R.}~\bibnamefont {Somma}}, \bibinfo {author}
  {\bibfnamefont {D.}~\bibnamefont {Gross}}, \bibinfo {author} {\bibfnamefont
  {S.~D.}\ \bibnamefont {Bartlett}}, \bibinfo {author} {\bibfnamefont
  {O.}~\bibnamefont {Landon-Cardinal}}, \bibinfo {author} {\bibfnamefont
  {D.}~\bibnamefont {Poulin}},\ and\ \bibinfo {author} {\bibfnamefont {Y.-K.}\
  \bibnamefont {Liu}},\ }\bibfield  {title} {\bibinfo {title} {Efficient
  quantum state tomography},\ }\href@noop {} {\bibfield  {journal} {\bibinfo
  {journal} {Nature Communications}\ }\textbf {\bibinfo {volume} {1}},\
  \bibinfo {pages} {1} (\bibinfo {year} {2010})}\BibitemShut {NoStop}%
\bibitem [{\citenamefont {Torlai}\ \emph {et~al.}(2018)\citenamefont {Torlai},
  \citenamefont {Mazzola}, \citenamefont {Carrasquilla}, \citenamefont
  {Troyer}, \citenamefont {Melko},\ and\ \citenamefont
  {Carleo}}]{torlai2018neural}%
  \BibitemOpen
  \bibfield  {author} {\bibinfo {author} {\bibfnamefont {G.}~\bibnamefont
  {Torlai}}, \bibinfo {author} {\bibfnamefont {G.}~\bibnamefont {Mazzola}},
  \bibinfo {author} {\bibfnamefont {J.}~\bibnamefont {Carrasquilla}}, \bibinfo
  {author} {\bibfnamefont {M.}~\bibnamefont {Troyer}}, \bibinfo {author}
  {\bibfnamefont {R.}~\bibnamefont {Melko}},\ and\ \bibinfo {author}
  {\bibfnamefont {G.}~\bibnamefont {Carleo}},\ }\bibfield  {title} {\bibinfo
  {title} {Neural-network quantum state tomography},\ }\href@noop {} {\bibfield
   {journal} {\bibinfo  {journal} {Nature Physics}\ }\textbf {\bibinfo {volume}
  {14}},\ \bibinfo {pages} {447} (\bibinfo {year} {2018})}\BibitemShut
  {NoStop}%
\bibitem [{\citenamefont {Orell}\ \emph {et~al.}(2019)\citenamefont {Orell},
  \citenamefont {Michailidis}, \citenamefont {Serbyn},\ and\ \citenamefont
  {Silveri}}]{Orell2019}%
  \BibitemOpen
  \bibfield  {author} {\bibinfo {author} {\bibfnamefont {T.}~\bibnamefont
  {Orell}}, \bibinfo {author} {\bibfnamefont {A.~A.}\ \bibnamefont
  {Michailidis}}, \bibinfo {author} {\bibfnamefont {M.}~\bibnamefont
  {Serbyn}},\ and\ \bibinfo {author} {\bibfnamefont {M.}~\bibnamefont
  {Silveri}},\ }\bibfield  {title} {\bibinfo {title} {Probing the many-body
  localization phase transition with superconducting circuits},\ }\href
  {https://doi.org/10.1103/PhysRevB.100.134504} {\bibfield  {journal} {\bibinfo
   {journal} {Physical Review B}\ }\textbf {\bibinfo {volume} {100}},\ \bibinfo
  {pages} {134504} (\bibinfo {year} {2019})}\BibitemShut {NoStop}%
\bibitem [{see()}]{seesm}%
  \BibitemOpen
  \bibfield  {title} {\bibinfo {title} {See supplementary materials, which
  includes refs.\cite{yan2019strongly,Wang2021,arute2019quantum, Aaronson2017,
  RevModPhys.91.021001,mori2018thermalization,Manai2015,White2020,sierant2019level,sierant2020model,atas2013distribution,oganesyan2007localization,cerezo2021cost,luitz2015many}},\
  }\href@noop {} {\ }\BibitemShut {NoStop}%
\bibitem [{\citenamefont {Kuhn}\ \emph {et~al.}(2007)\citenamefont {Kuhn},
  \citenamefont {Sigwarth}, \citenamefont {Miniatura}, \citenamefont
  {Delande},\ and\ \citenamefont {Müller}}]{Kuhn_2007}%
  \BibitemOpen
  \bibfield  {author} {\bibinfo {author} {\bibfnamefont {R.~C.}\ \bibnamefont
  {Kuhn}}, \bibinfo {author} {\bibfnamefont {O.}~\bibnamefont {Sigwarth}},
  \bibinfo {author} {\bibfnamefont {C.}~\bibnamefont {Miniatura}}, \bibinfo
  {author} {\bibfnamefont {D.}~\bibnamefont {Delande}},\ and\ \bibinfo {author}
  {\bibfnamefont {C.~A.}\ \bibnamefont {Müller}},\ }\bibfield  {title}
  {\bibinfo {title} {Coherent matter wave transport in speckle potentials},\
  }\href {https://doi.org/10.1088/1367-2630/9/6/161} {\bibfield  {journal}
  {\bibinfo  {journal} {New Journal of Physics}\ }\textbf {\bibinfo {volume}
  {9}},\ \bibinfo {pages} {161} (\bibinfo {year} {2007})}\BibitemShut {NoStop}%
\bibitem [{\citenamefont {Escalante}\ and\ \citenamefont
  {Skipetrov}(2018)}]{Escalante2018}%
  \BibitemOpen
  \bibfield  {author} {\bibinfo {author} {\bibfnamefont {J.~M.}\ \bibnamefont
  {Escalante}}\ and\ \bibinfo {author} {\bibfnamefont {S.~E.}\ \bibnamefont
  {Skipetrov}},\ }\bibfield  {title} {\bibinfo {title} {{Level spacing
  statistics for light in two-dimensional disordered photonic crystals}},\
  }\href {https://doi.org/10.1038/s41598-018-29996-1} {\bibfield  {journal}
  {\bibinfo  {journal} {Scientific Reports}\ }\textbf {\bibinfo {volume} {8}},\
  \bibinfo {pages} {11569} (\bibinfo {year} {2018})}\BibitemShut {NoStop}%
\bibitem [{\citenamefont {Zhang}\ \emph {et~al.}(2021)\citenamefont {Zhang},
  \citenamefont {Wang}, \citenamefont {Zheng}, \citenamefont {Sun},
  \citenamefont {Sun}, \citenamefont {Wang}, \citenamefont {Schirmacher},\ and\
  \citenamefont {Zhang}}]{Zhang2021}%
  \BibitemOpen
  \bibfield  {author} {\bibinfo {author} {\bibfnamefont {L.}~\bibnamefont
  {Zhang}}, \bibinfo {author} {\bibfnamefont {Y.}~\bibnamefont {Wang}},
  \bibinfo {author} {\bibfnamefont {J.}~\bibnamefont {Zheng}}, \bibinfo
  {author} {\bibfnamefont {A.}~\bibnamefont {Sun}}, \bibinfo {author}
  {\bibfnamefont {X.}~\bibnamefont {Sun}}, \bibinfo {author} {\bibfnamefont
  {Y.}~\bibnamefont {Wang}}, \bibinfo {author} {\bibfnamefont {W.}~\bibnamefont
  {Schirmacher}},\ and\ \bibinfo {author} {\bibfnamefont {J.}~\bibnamefont
  {Zhang}},\ }\bibfield  {title} {\bibinfo {title} {{Level statistics and
  Anderson delocalization in two-dimensional granular materials}},\ }\href
  {https://doi.org/10.1103/PhysRevB.103.104201} {\bibfield  {journal} {\bibinfo
   {journal} {Physical Review B}\ }\textbf {\bibinfo {volume} {103}},\ \bibinfo
  {pages} {104201} (\bibinfo {year} {2021})}\BibitemShut {NoStop}%
\bibitem [{\citenamefont {Cong}\ \emph {et~al.}(2019)\citenamefont {Cong},
  \citenamefont {Choi},\ and\ \citenamefont {Lukin}}]{cong2019quantum}%
  \BibitemOpen
  \bibfield  {author} {\bibinfo {author} {\bibfnamefont {I.}~\bibnamefont
  {Cong}}, \bibinfo {author} {\bibfnamefont {S.}~\bibnamefont {Choi}},\ and\
  \bibinfo {author} {\bibfnamefont {M.~D.}\ \bibnamefont {Lukin}},\ }\bibfield
  {title} {\bibinfo {title} {Quantum convolutional neural networks},\
  }\href@noop {} {\bibfield  {journal} {\bibinfo  {journal} {Nature Physics}\
  }\textbf {\bibinfo {volume} {15}},\ \bibinfo {pages} {1273} (\bibinfo {year}
  {2019})}\BibitemShut {NoStop}%
\bibitem [{\citenamefont {Liu}\ \emph {et~al.}(2021)\citenamefont {Liu},
  \citenamefont {Lim}, \citenamefont {Wood}, \citenamefont {Huang},
  \citenamefont {Guo},\ and\ \citenamefont {Huang}}]{liuqccnn2021}%
  \BibitemOpen
  \bibfield  {author} {\bibinfo {author} {\bibfnamefont {J.}~\bibnamefont
  {Liu}}, \bibinfo {author} {\bibfnamefont {K.~H.}\ \bibnamefont {Lim}},
  \bibinfo {author} {\bibfnamefont {K.~L.}\ \bibnamefont {Wood}}, \bibinfo
  {author} {\bibfnamefont {W.}~\bibnamefont {Huang}}, \bibinfo {author}
  {\bibfnamefont {C.}~\bibnamefont {Guo}},\ and\ \bibinfo {author}
  {\bibfnamefont {H.-L.}\ \bibnamefont {Huang}},\ }\bibfield  {title} {\bibinfo
  {title} {Hybrid quantum-classical convolutional neural networks},\
  }\href@noop {} {\bibfield  {journal} {\bibinfo  {journal} {Science China
  Physics, Mechanics \& Astronomy}\ }\textbf {\bibinfo {volume} {64}},\
  \bibinfo {pages} {290311} (\bibinfo {year} {2021})}\BibitemShut {NoStop}%
\bibitem [{\citenamefont {Benedetti}\ \emph {et~al.}(2019)\citenamefont
  {Benedetti}, \citenamefont {Lloyd}, \citenamefont {Sack},\ and\ \citenamefont
  {Fiorentini}}]{benedetti2019parameterized}%
  \BibitemOpen
  \bibfield  {author} {\bibinfo {author} {\bibfnamefont {M.}~\bibnamefont
  {Benedetti}}, \bibinfo {author} {\bibfnamefont {E.}~\bibnamefont {Lloyd}},
  \bibinfo {author} {\bibfnamefont {S.}~\bibnamefont {Sack}},\ and\ \bibinfo
  {author} {\bibfnamefont {M.}~\bibnamefont {Fiorentini}},\ }\bibfield  {title}
  {\bibinfo {title} {Parameterized quantum circuits as machine learning
  models},\ }\href@noop {} {\bibfield  {journal} {\bibinfo  {journal} {Quantum
  Science and Technology}\ }\textbf {\bibinfo {volume} {4}},\ \bibinfo {pages}
  {043001} (\bibinfo {year} {2019})}\BibitemShut {NoStop}%
\bibitem [{\citenamefont {Mitarai}\ \emph {et~al.}(2018)\citenamefont
  {Mitarai}, \citenamefont {Negoro}, \citenamefont {Kitagawa},\ and\
  \citenamefont {Fujii}}]{mitarai2018quantum}%
  \BibitemOpen
  \bibfield  {author} {\bibinfo {author} {\bibfnamefont {K.}~\bibnamefont
  {Mitarai}}, \bibinfo {author} {\bibfnamefont {M.}~\bibnamefont {Negoro}},
  \bibinfo {author} {\bibfnamefont {M.}~\bibnamefont {Kitagawa}},\ and\
  \bibinfo {author} {\bibfnamefont {K.}~\bibnamefont {Fujii}},\ }\bibfield
  {title} {\bibinfo {title} {Quantum circuit learning},\ }\href@noop {}
  {\bibfield  {journal} {\bibinfo  {journal} {Physical Review A}\ }\textbf
  {\bibinfo {volume} {98}},\ \bibinfo {pages} {032309} (\bibinfo {year}
  {2018})}\BibitemShut {NoStop}%
\bibitem [{\citenamefont {Ruby}\ and\ \citenamefont
  {Yendapalli}(2020)}]{ruby2020binary}%
  \BibitemOpen
  \bibfield  {author} {\bibinfo {author} {\bibfnamefont {U.}~\bibnamefont
  {Ruby}}\ and\ \bibinfo {author} {\bibfnamefont {V.}~\bibnamefont
  {Yendapalli}},\ }\bibfield  {title} {\bibinfo {title} {Binary cross entropy
  with deep learning technique for image classification},\ }\href@noop {}
  {\bibfield  {journal} {\bibinfo  {journal} {Int. J. Adv. Trends Comput. Sci.
  Eng}\ }\textbf {\bibinfo {volume} {9}} (\bibinfo {year} {2020})}\BibitemShut
  {NoStop}%
\bibitem [{\citenamefont {Mannor}\ \emph {et~al.}(2005)\citenamefont {Mannor},
  \citenamefont {Peleg},\ and\ \citenamefont {Rubinstein}}]{mannor2005cross}%
  \BibitemOpen
  \bibfield  {author} {\bibinfo {author} {\bibfnamefont {S.}~\bibnamefont
  {Mannor}}, \bibinfo {author} {\bibfnamefont {D.}~\bibnamefont {Peleg}},\ and\
  \bibinfo {author} {\bibfnamefont {R.}~\bibnamefont {Rubinstein}},\ }\bibfield
   {title} {\bibinfo {title} {The cross entropy method for classification},\
  }in\ \href@noop {} {\emph {\bibinfo {booktitle} {Proceedings of the 22nd
  international conference on Machine learning}}}\ (\bibinfo {year} {2005})\
  pp.\ \bibinfo {pages} {561--568}\BibitemShut {NoStop}%
\bibitem [{\citenamefont {Ho}\ and\ \citenamefont {Wookey}(2019)}]{ho2019real}%
  \BibitemOpen
  \bibfield  {author} {\bibinfo {author} {\bibfnamefont {Y.}~\bibnamefont
  {Ho}}\ and\ \bibinfo {author} {\bibfnamefont {S.}~\bibnamefont {Wookey}},\
  }\bibfield  {title} {\bibinfo {title} {The real-world-weight cross-entropy
  loss function: Modeling the costs of mislabeling},\ }\href@noop {} {\bibfield
   {journal} {\bibinfo  {journal} {IEEE Access}\ }\textbf {\bibinfo {volume}
  {8}},\ \bibinfo {pages} {4806} (\bibinfo {year} {2019})}\BibitemShut
  {NoStop}%
\bibitem [{\citenamefont {Schuld}\ \emph {et~al.}(2019)\citenamefont {Schuld},
  \citenamefont {Bergholm}, \citenamefont {Gogolin}, \citenamefont {Izaac},\
  and\ \citenamefont {Killoran}}]{schuld2019evaluating}%
  \BibitemOpen
  \bibfield  {author} {\bibinfo {author} {\bibfnamefont {M.}~\bibnamefont
  {Schuld}}, \bibinfo {author} {\bibfnamefont {V.}~\bibnamefont {Bergholm}},
  \bibinfo {author} {\bibfnamefont {C.}~\bibnamefont {Gogolin}}, \bibinfo
  {author} {\bibfnamefont {J.}~\bibnamefont {Izaac}},\ and\ \bibinfo {author}
  {\bibfnamefont {N.}~\bibnamefont {Killoran}},\ }\bibfield  {title} {\bibinfo
  {title} {Evaluating analytic gradients on quantum hardware},\ }\href@noop {}
  {\bibfield  {journal} {\bibinfo  {journal} {Physical Review A}\ }\textbf
  {\bibinfo {volume} {99}},\ \bibinfo {pages} {032331} (\bibinfo {year}
  {2019})}\BibitemShut {NoStop}%
\bibitem [{\citenamefont {Parra-Rodriguez}\ \emph {et~al.}(2020)\citenamefont
  {Parra-Rodriguez}, \citenamefont {Lougovski}, \citenamefont {Lamata},
  \citenamefont {Solano},\ and\ \citenamefont {Sanz}}]{ParraDigital2020}%
  \BibitemOpen
  \bibfield  {author} {\bibinfo {author} {\bibfnamefont {A.}~\bibnamefont
  {Parra-Rodriguez}}, \bibinfo {author} {\bibfnamefont {P.}~\bibnamefont
  {Lougovski}}, \bibinfo {author} {\bibfnamefont {L.}~\bibnamefont {Lamata}},
  \bibinfo {author} {\bibfnamefont {E.}~\bibnamefont {Solano}},\ and\ \bibinfo
  {author} {\bibfnamefont {M.}~\bibnamefont {Sanz}},\ }\bibfield  {title}
  {\bibinfo {title} {Digital-analog quantum computation},\ }\href
  {https://doi.org/10.1103/PhysRevA.101.022305} {\bibfield  {journal} {\bibinfo
   {journal} {Physical Review A}\ }\textbf {\bibinfo {volume} {101}},\ \bibinfo
  {pages} {022305} (\bibinfo {year} {2020})}\BibitemShut {NoStop}%
\bibitem [{\citenamefont {Yan}\ \emph {et~al.}(2019)\citenamefont {Yan},
  \citenamefont {Zhang}, \citenamefont {Gong}, \citenamefont {Wu},
  \citenamefont {Zheng}, \citenamefont {Li}, \citenamefont {Wang},
  \citenamefont {Liang}, \citenamefont {Lin}, \citenamefont {Xu} \emph
  {et~al.}}]{yan2019strongly}%
  \BibitemOpen
  \bibfield  {author} {\bibinfo {author} {\bibfnamefont {Z.}~\bibnamefont
  {Yan}}, \bibinfo {author} {\bibfnamefont {Y.-R.}\ \bibnamefont {Zhang}},
  \bibinfo {author} {\bibfnamefont {M.}~\bibnamefont {Gong}}, \bibinfo {author}
  {\bibfnamefont {Y.}~\bibnamefont {Wu}}, \bibinfo {author} {\bibfnamefont
  {Y.}~\bibnamefont {Zheng}}, \bibinfo {author} {\bibfnamefont
  {S.}~\bibnamefont {Li}}, \bibinfo {author} {\bibfnamefont {C.}~\bibnamefont
  {Wang}}, \bibinfo {author} {\bibfnamefont {F.}~\bibnamefont {Liang}},
  \bibinfo {author} {\bibfnamefont {J.}~\bibnamefont {Lin}}, \bibinfo {author}
  {\bibfnamefont {Y.}~\bibnamefont {Xu}}, \emph {et~al.},\ }\bibfield  {title}
  {\bibinfo {title} {Strongly correlated quantum walks with a 12-qubit
  superconducting processor},\ }\href@noop {} {\bibfield  {journal} {\bibinfo
  {journal} {Science}\ }\textbf {\bibinfo {volume} {364}},\ \bibinfo {pages}
  {753} (\bibinfo {year} {2019})}\BibitemShut {NoStop}%
\bibitem [{\citenamefont {Wang}\ \emph {et~al.}(2021)\citenamefont {Wang},
  \citenamefont {Chen}, \citenamefont {Lu},\ and\ \citenamefont
  {Pan}}]{Wang2021}%
  \BibitemOpen
  \bibfield  {author} {\bibinfo {author} {\bibfnamefont {C.}~\bibnamefont
  {Wang}}, \bibinfo {author} {\bibfnamefont {M.-C.}\ \bibnamefont {Chen}},
  \bibinfo {author} {\bibfnamefont {C.-Y.}\ \bibnamefont {Lu}},\ and\ \bibinfo
  {author} {\bibfnamefont {J.-W.}\ \bibnamefont {Pan}},\ }\bibfield  {title}
  {\bibinfo {title} {{Optimal readout of superconducting qubits exploiting
  high-level states}},\ }\href {https://doi.org/10.1016/j.fmre.2020.12.008}
  {\bibfield  {journal} {\bibinfo  {journal} {Fundamental Research}\ }\textbf
  {\bibinfo {volume} {1}},\ \bibinfo {pages} {16} (\bibinfo {year}
  {2021})}\BibitemShut {NoStop}%
\bibitem [{\citenamefont {Arute}\ \emph {et~al.}(2019)\citenamefont {Arute},
  \citenamefont {Arya}, \citenamefont {Babbush}, \citenamefont {Bacon},
  \citenamefont {Bardin}, \citenamefont {Barends}, \citenamefont {Biswas},
  \citenamefont {Boixo}, \citenamefont {Brandao}, \citenamefont {Buell} \emph
  {et~al.}}]{arute2019quantum}%
  \BibitemOpen
  \bibfield  {author} {\bibinfo {author} {\bibfnamefont {F.}~\bibnamefont
  {Arute}}, \bibinfo {author} {\bibfnamefont {K.}~\bibnamefont {Arya}},
  \bibinfo {author} {\bibfnamefont {R.}~\bibnamefont {Babbush}}, \bibinfo
  {author} {\bibfnamefont {D.}~\bibnamefont {Bacon}}, \bibinfo {author}
  {\bibfnamefont {J.~C.}\ \bibnamefont {Bardin}}, \bibinfo {author}
  {\bibfnamefont {R.}~\bibnamefont {Barends}}, \bibinfo {author} {\bibfnamefont
  {R.}~\bibnamefont {Biswas}}, \bibinfo {author} {\bibfnamefont
  {S.}~\bibnamefont {Boixo}}, \bibinfo {author} {\bibfnamefont {F.~G.}\
  \bibnamefont {Brandao}}, \bibinfo {author} {\bibfnamefont {D.~A.}\
  \bibnamefont {Buell}}, \emph {et~al.},\ }\bibfield  {title} {\bibinfo {title}
  {Quantum supremacy using a programmable superconducting processor},\
  }\href@noop {} {\bibfield  {journal} {\bibinfo  {journal} {Nature}\ }\textbf
  {\bibinfo {volume} {574}},\ \bibinfo {pages} {505} (\bibinfo {year}
  {2019})}\BibitemShut {NoStop}%
\bibitem [{\citenamefont {Aaronson}\ and\ \citenamefont
  {Chen}(2017)}]{Aaronson2017}%
  \BibitemOpen
  \bibfield  {author} {\bibinfo {author} {\bibfnamefont {S.}~\bibnamefont
  {Aaronson}}\ and\ \bibinfo {author} {\bibfnamefont {L.}~\bibnamefont
  {Chen}},\ }\bibfield  {title} {\bibinfo {title} {{Complexity-theoretic
  foundations of quantum supremacy experiments}},\ }in\ \href
  {https://doi.org/10.4230/LIPIcs.CCC.2017.22} {\emph {\bibinfo {booktitle}
  {Leibniz International Proceedings in Informatics, LIPIcs}}},\ Vol.~\bibinfo
  {volume} {79}\ (\bibinfo  {publisher} {Schloss Dagstuhl- Leibniz-Zentrum fur
  Informatik GmbH, Dagstuhl Publishing},\ \bibinfo {year} {2017})\BibitemShut
  {NoStop}%
\bibitem [{\citenamefont {Abanin}\ \emph {et~al.}(2019)\citenamefont {Abanin},
  \citenamefont {Altman}, \citenamefont {Bloch},\ and\ \citenamefont
  {Serbyn}}]{RevModPhys.91.021001}%
  \BibitemOpen
  \bibfield  {author} {\bibinfo {author} {\bibfnamefont {D.~A.}\ \bibnamefont
  {Abanin}}, \bibinfo {author} {\bibfnamefont {E.}~\bibnamefont {Altman}},
  \bibinfo {author} {\bibfnamefont {I.}~\bibnamefont {Bloch}},\ and\ \bibinfo
  {author} {\bibfnamefont {M.}~\bibnamefont {Serbyn}},\ }\bibfield  {title}
  {\bibinfo {title} {Colloquium: Many-body localization, thermalization, and
  entanglement},\ }\href {https://doi.org/10.1103/RevModPhys.91.021001}
  {\bibfield  {journal} {\bibinfo  {journal} {Reviews of Modern Physics}\
  }\textbf {\bibinfo {volume} {91}},\ \bibinfo {pages} {021001} (\bibinfo
  {year} {2019})}\BibitemShut {NoStop}%
\bibitem [{\citenamefont {Mori}\ \emph {et~al.}(2018)\citenamefont {Mori},
  \citenamefont {Ikeda}, \citenamefont {Kaminishi},\ and\ \citenamefont
  {Ueda}}]{mori2018thermalization}%
  \BibitemOpen
  \bibfield  {author} {\bibinfo {author} {\bibfnamefont {T.}~\bibnamefont
  {Mori}}, \bibinfo {author} {\bibfnamefont {T.~N.}\ \bibnamefont {Ikeda}},
  \bibinfo {author} {\bibfnamefont {E.}~\bibnamefont {Kaminishi}},\ and\
  \bibinfo {author} {\bibfnamefont {M.}~\bibnamefont {Ueda}},\ }\bibfield
  {title} {\bibinfo {title} {Thermalization and prethermalization in isolated
  quantum systems: a theoretical overview},\ }\href@noop {} {\bibfield
  {journal} {\bibinfo  {journal} {Journal of Physics B: Atomic, Molecular and
  Optical Physics}\ }\textbf {\bibinfo {volume} {51}},\ \bibinfo {pages}
  {112001} (\bibinfo {year} {2018})}\BibitemShut {NoStop}%
\bibitem [{\citenamefont {Manai}\ \emph {et~al.}(2015)\citenamefont {Manai},
  \citenamefont {Cl{\'{e}}ment}, \citenamefont {Chicireanu}, \citenamefont
  {Hainaut}, \citenamefont {Garreau}, \citenamefont {Szriftgiser},\ and\
  \citenamefont {Delande}}]{Manai2015}%
  \BibitemOpen
  \bibfield  {author} {\bibinfo {author} {\bibfnamefont {I.}~\bibnamefont
  {Manai}}, \bibinfo {author} {\bibfnamefont {J.-F.}\ \bibnamefont
  {Cl{\'{e}}ment}}, \bibinfo {author} {\bibfnamefont {R.}~\bibnamefont
  {Chicireanu}}, \bibinfo {author} {\bibfnamefont {C.}~\bibnamefont {Hainaut}},
  \bibinfo {author} {\bibfnamefont {J.~C.}\ \bibnamefont {Garreau}}, \bibinfo
  {author} {\bibfnamefont {P.}~\bibnamefont {Szriftgiser}},\ and\ \bibinfo
  {author} {\bibfnamefont {D.}~\bibnamefont {Delande}},\ }\bibfield  {title}
  {\bibinfo {title} {{Experimental Observation of Two-Dimensional Anderson
  Localization with the Atomic Kicked Rotor}},\ }\href
  {https://doi.org/10.1103/PhysRevLett.115.240603} {\bibfield  {journal}
  {\bibinfo  {journal} {Physical Review Letters}\ }\textbf {\bibinfo {volume}
  {115}},\ \bibinfo {pages} {240603} (\bibinfo {year} {2015})}\BibitemShut
  {NoStop}%
\bibitem [{\citenamefont {White}\ \emph {et~al.}(2020)\citenamefont {White},
  \citenamefont {Haase}, \citenamefont {Brown}, \citenamefont {Hoogerland},
  \citenamefont {Najafabadi}, \citenamefont {Helm}, \citenamefont {Gies},
  \citenamefont {Schumayer},\ and\ \citenamefont {Hutchinson}}]{White2020}%
  \BibitemOpen
  \bibfield  {author} {\bibinfo {author} {\bibfnamefont {D.~H.}\ \bibnamefont
  {White}}, \bibinfo {author} {\bibfnamefont {T.~A.}\ \bibnamefont {Haase}},
  \bibinfo {author} {\bibfnamefont {D.~J.}\ \bibnamefont {Brown}}, \bibinfo
  {author} {\bibfnamefont {M.~D.}\ \bibnamefont {Hoogerland}}, \bibinfo
  {author} {\bibfnamefont {M.~S.}\ \bibnamefont {Najafabadi}}, \bibinfo
  {author} {\bibfnamefont {J.~L.}\ \bibnamefont {Helm}}, \bibinfo {author}
  {\bibfnamefont {C.}~\bibnamefont {Gies}}, \bibinfo {author} {\bibfnamefont
  {D.}~\bibnamefont {Schumayer}},\ and\ \bibinfo {author} {\bibfnamefont
  {D.~A.~W.}\ \bibnamefont {Hutchinson}},\ }\bibfield  {title} {\bibinfo
  {title} {{Observation of two-dimensional Anderson localisation of ultracold
  atoms}},\ }\href {https://doi.org/10.1038/s41467-020-18652-w} {\bibfield
  {journal} {\bibinfo  {journal} {Nature Communications}\ }\textbf {\bibinfo
  {volume} {11}},\ \bibinfo {pages} {4942} (\bibinfo {year}
  {2020})}\BibitemShut {NoStop}%
\bibitem [{\citenamefont {Sierant}\ and\ \citenamefont
  {Zakrzewski}(2019)}]{sierant2019level}%
  \BibitemOpen
  \bibfield  {author} {\bibinfo {author} {\bibfnamefont {P.}~\bibnamefont
  {Sierant}}\ and\ \bibinfo {author} {\bibfnamefont {J.}~\bibnamefont
  {Zakrzewski}},\ }\bibfield  {title} {\bibinfo {title} {Level statistics
  across the many-body localization transition},\ }\href@noop {} {\bibfield
  {journal} {\bibinfo  {journal} {Physical Review B}\ }\textbf {\bibinfo
  {volume} {99}},\ \bibinfo {pages} {104205} (\bibinfo {year}
  {2019})}\BibitemShut {NoStop}%
\bibitem [{\citenamefont {Sierant}\ and\ \citenamefont
  {Zakrzewski}(2020)}]{sierant2020model}%
  \BibitemOpen
  \bibfield  {author} {\bibinfo {author} {\bibfnamefont {P.}~\bibnamefont
  {Sierant}}\ and\ \bibinfo {author} {\bibfnamefont {J.}~\bibnamefont
  {Zakrzewski}},\ }\bibfield  {title} {\bibinfo {title} {Model of level
  statistics for disordered interacting quantum many-body systems},\
  }\href@noop {} {\bibfield  {journal} {\bibinfo  {journal} {Physical Review
  B}\ }\textbf {\bibinfo {volume} {101}},\ \bibinfo {pages} {104201} (\bibinfo
  {year} {2020})}\BibitemShut {NoStop}%
\bibitem [{\citenamefont {Atas}\ \emph {et~al.}(2013)\citenamefont {Atas},
  \citenamefont {Bogomolny}, \citenamefont {Giraud},\ and\ \citenamefont
  {Roux}}]{atas2013distribution}%
  \BibitemOpen
  \bibfield  {author} {\bibinfo {author} {\bibfnamefont {Y.}~\bibnamefont
  {Atas}}, \bibinfo {author} {\bibfnamefont {E.}~\bibnamefont {Bogomolny}},
  \bibinfo {author} {\bibfnamefont {O.}~\bibnamefont {Giraud}},\ and\ \bibinfo
  {author} {\bibfnamefont {G.}~\bibnamefont {Roux}},\ }\bibfield  {title}
  {\bibinfo {title} {Distribution of the ratio of consecutive level spacings in
  random matrix ensembles},\ }\href@noop {} {\bibfield  {journal} {\bibinfo
  {journal} {Physical Review Letters}\ }\textbf {\bibinfo {volume} {110}},\
  \bibinfo {pages} {084101} (\bibinfo {year} {2013})}\BibitemShut {NoStop}%
\bibitem [{\citenamefont {Oganesyan}\ and\ \citenamefont
  {Huse}(2007)}]{oganesyan2007localization}%
  \BibitemOpen
  \bibfield  {author} {\bibinfo {author} {\bibfnamefont {V.}~\bibnamefont
  {Oganesyan}}\ and\ \bibinfo {author} {\bibfnamefont {D.~A.}\ \bibnamefont
  {Huse}},\ }\bibfield  {title} {\bibinfo {title} {Localization of interacting
  fermions at high temperature},\ }\href@noop {} {\bibfield  {journal}
  {\bibinfo  {journal} {Physical Review B}\ }\textbf {\bibinfo {volume} {75}},\
  \bibinfo {pages} {155111} (\bibinfo {year} {2007})}\BibitemShut {NoStop}%
\bibitem [{\citenamefont {Cerezo}\ \emph {et~al.}(2021)\citenamefont {Cerezo},
  \citenamefont {Sone}, \citenamefont {Volkoff}, \citenamefont {Cincio},\ and\
  \citenamefont {Coles}}]{cerezo2021cost}%
  \BibitemOpen
  \bibfield  {author} {\bibinfo {author} {\bibfnamefont {M.}~\bibnamefont
  {Cerezo}}, \bibinfo {author} {\bibfnamefont {A.}~\bibnamefont {Sone}},
  \bibinfo {author} {\bibfnamefont {T.}~\bibnamefont {Volkoff}}, \bibinfo
  {author} {\bibfnamefont {L.}~\bibnamefont {Cincio}},\ and\ \bibinfo {author}
  {\bibfnamefont {P.~J.}\ \bibnamefont {Coles}},\ }\bibfield  {title} {\bibinfo
  {title} {Cost function dependent barren plateaus in shallow parametrized
  quantum circuits},\ }\href@noop {} {\bibfield  {journal} {\bibinfo  {journal}
  {Nature communications}\ }\textbf {\bibinfo {volume} {12}},\ \bibinfo {pages}
  {1} (\bibinfo {year} {2021})}\BibitemShut {NoStop}%
\bibitem [{\citenamefont {Luitz}\ \emph {et~al.}(2015)\citenamefont {Luitz},
  \citenamefont {Laflorencie},\ and\ \citenamefont {Alet}}]{luitz2015many}%
  \BibitemOpen
  \bibfield  {author} {\bibinfo {author} {\bibfnamefont {D.~J.}\ \bibnamefont
  {Luitz}}, \bibinfo {author} {\bibfnamefont {N.}~\bibnamefont {Laflorencie}},\
  and\ \bibinfo {author} {\bibfnamefont {F.}~\bibnamefont {Alet}},\ }\bibfield
  {title} {\bibinfo {title} {Many-body localization edge in the random-field
  heisenberg chain},\ }\href@noop {} {\bibfield  {journal} {\bibinfo  {journal}
  {Physical Review B}\ }\textbf {\bibinfo {volume} {91}},\ \bibinfo {pages}
  {081103} (\bibinfo {year} {2015})}\BibitemShut {NoStop}%
\end{thebibliography}%


%

\end{document}


\title{Supplementary Materials for ``Quantum Neuronal Sensing of Quantum Many-Body States on a 61-Qubit Programmable Superconducting Processor''}

\author{Ming Gong$^{1,2,3}$}
\thanks{These authors contributed equally to this work.}
\author{He-Liang Huang$^{1,2,3}$}
\thanks{These authors contributed equally to this work.}
\author{Shiyu Wang$^{1,2,3}$}
\thanks{These authors contributed equally to this work.}
\author{Chu Guo$^{1,2,3}$}
\author{Shaowei Li$^{1,2,3}$}
\author{Yulin Wu$^{1,2,3}$}
\author{Qingling Zhu$^{1,2,3}$}
\author{Youwei Zhao$^{1,2,3}$}
\author{Shaojun Guo$^{1,2,3}$}
\author{Haoran Qian$^{1,2,3}$}
\author{Yangsen Ye$^{1,2,3}$}
\author{Chen Zha$^{1,2,3}$}
\author{Fusheng Chen$^{1,2,3}$}
\author{Chong Ying$^{1,2,3}$}
\author{Jiale Yu$^{1,2,3}$}
\author{Daojin Fan$^{1,2,3}$}
\author{Dachao Wu$^{1,2,3}$}
\author{Hong Su$^{1,2,3}$}
\author{Hui Deng$^{1,2,3}$}
\author{Hao Rong$^{1,2,3}$}
\author{Kaili Zhang$^{1,2,3}$}
\author{Sirui Cao$^{1,2,3}$}
\author{Jin Lin$^{1,2,3}$}
\author{Yu Xu$^{1,2,3}$}
\author{Lihua Sun$^{1,2,3}$}
\author{Cheng Guo$^{1,2,3}$}
\author{Na Li$^{1,2,3}$}
\author{Futian Liang$^{1,2,3}$}
\author{Akitada Sakurai$^{4,6}$}
\author{Kae Nemoto$^{6,7,4}$}
\author{W. J. Munro$^{5,6}$} \email{bill.munro@me.com}
\author{Yong-Heng Huo$^{1,2,3}$}
\author{Chao-Yang Lu$^{1,2,3}$}
\author{Cheng-Zhi Peng$^{1,2,3}$}
\author{Xiaobo Zhu$^{1,2,3}$} \email{xbzhu16@ustc.edu.cn}
\author{Jian-Wei Pan$^{1,2,3}$} \email{pan@ustc.edu.cn}

\affiliation{$^1$ Department of Modern Physics, University of Science and Technology of China, Hefei 230026, China}
\affiliation{$^2$  Shanghai Branch, CAS Center for Excellence in Quantum Information and Quantum Physics, University of Science and Technology of China, Shanghai 201315, China}
\affiliation{$^3$  Shanghai Research Center for Quantum Sciences, Shanghai 201315, China}
\affiliation{$^4$  School of Multidisciplinary Science, Department of Informatics, SOKENDAI (the Graduate University for Advanced Studies), 2-1-2 Hitotsubashi, Chiyoda-ku, Tokyo 101-8430, Japan}
\affiliation{$^5$  NTT Basic Research Laboratories and Research Center for Theoretical Quantum Physics, 3-1 Morinosato-Wakamiya, Atsugi, Kanagawa 243-0198, Japan}
\affiliation{$^6$  National Institute of Informatics, 2-1-2 Hitotsubashi, Chiyoda-ku, Tokyo 101-8430, Japan}
\affiliation{$^7$ Okinawa Institute of Science and Technology Graduate University, Onna-son, Okinawa 904-0495, Japan}

\maketitle
\tableofcontents

\beginsupplement
\section{Quantum processor performance}
\begin{figure*}[hbt]
\centering
\includegraphics[width=0.45\linewidth]{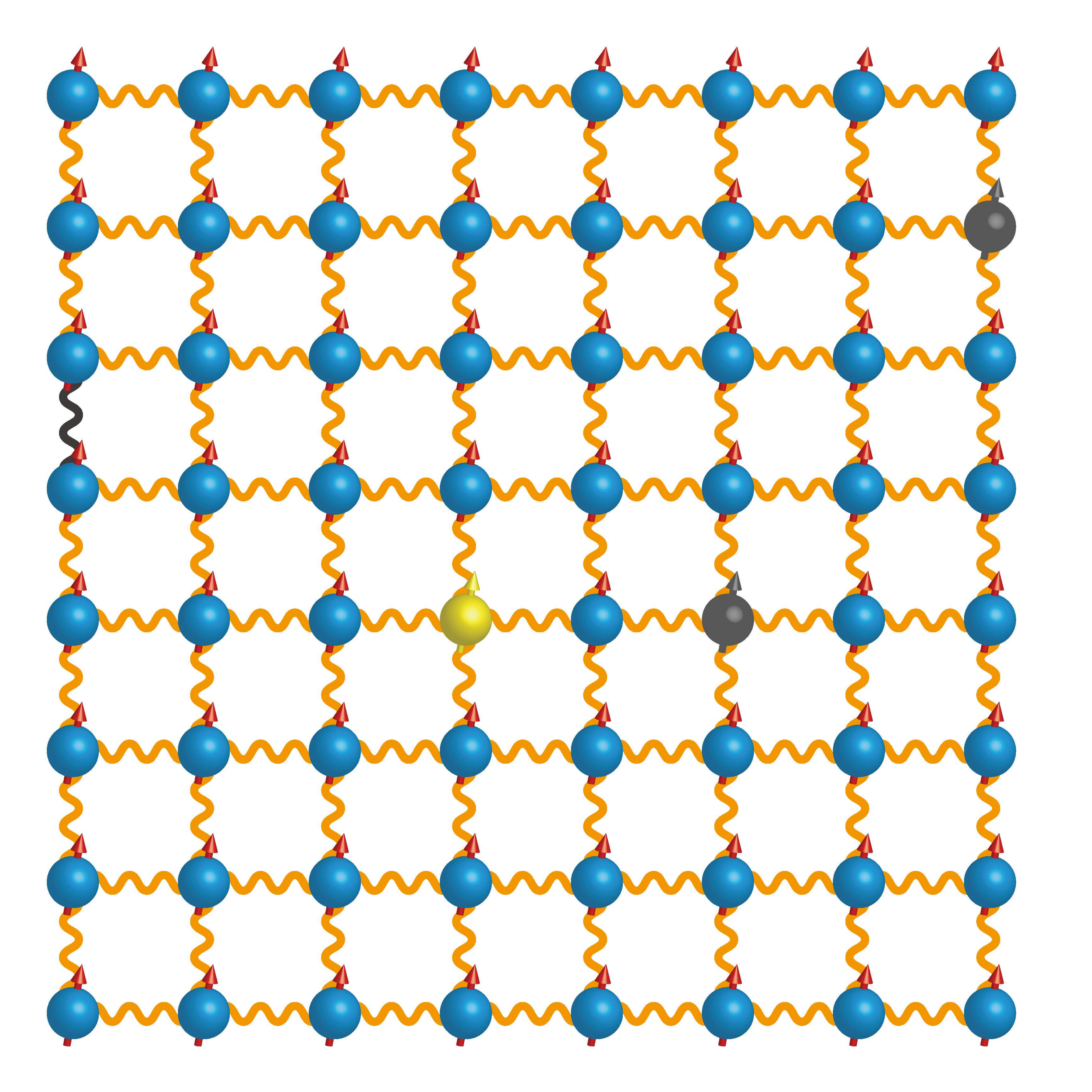}
\caption{\textbf{Sketch of the \textit{Zuchongzhi 1.0} quantum processor.} Each ball with an arrow represents a qubit while link between qubits represents the coupling resonators. The qubits and resonator colored dark gray are non functional while the qubit colored yellow are not fast-tunable.
}
\label{figS0}
\end{figure*}

The experiment is performed on a superconducting quantum computing platform as in [Science 372, 948-952 (2021)]~\cite{Gong2021}. The quantum processor, namely ``\textit{Zuchongzhi} 1.0'', contains 64 transmon qubits arranged in an 8-by-8 array, among them two qubits are not functional, and one qubit that is not frequency fast-tunable (Fig.~\ref{figS0}). Between neighboring qubits, a coplanar waveguide resonator couples both qubits. Through the virtual photon interaction via the quantum bus, an effective coupling between neighboring qubits is realized.

\blue{In realizing the high-fidelity continuous time Hamiltonian evolution, where $H_d=\hbar \sum g_{i,j}(\sigma_x^i\sigma_x^j+\sigma_y^i\sigma_y^j)/2+\hbar \sum d_i\sigma_z^i$, we need to realize a high-precision control of qubit frequency. In this work, we utilize two technologies to realize the requirement: the Z pulse distortion correction, and the alignment of qubit frequency.}

\begin{figure*}[hbt]
\centering \includegraphics[width=0.5\linewidth]{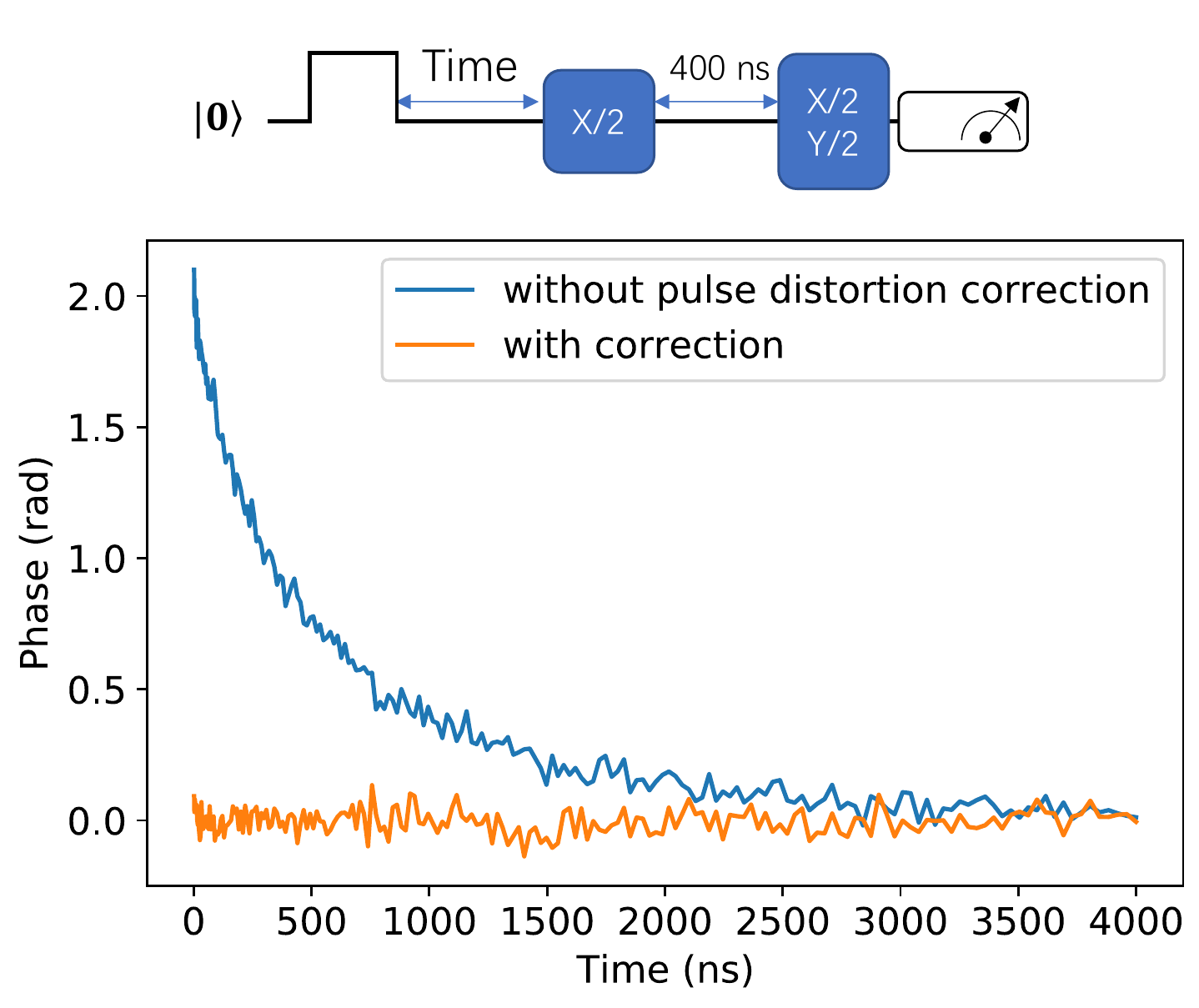}
\caption{\textbf{The wave sequence and results for pulse distortion correction.}
}
\label{pulsecorrection}
\end{figure*}

\begin{figure*}[hbt]
\centering \includegraphics[width=0.5\linewidth]{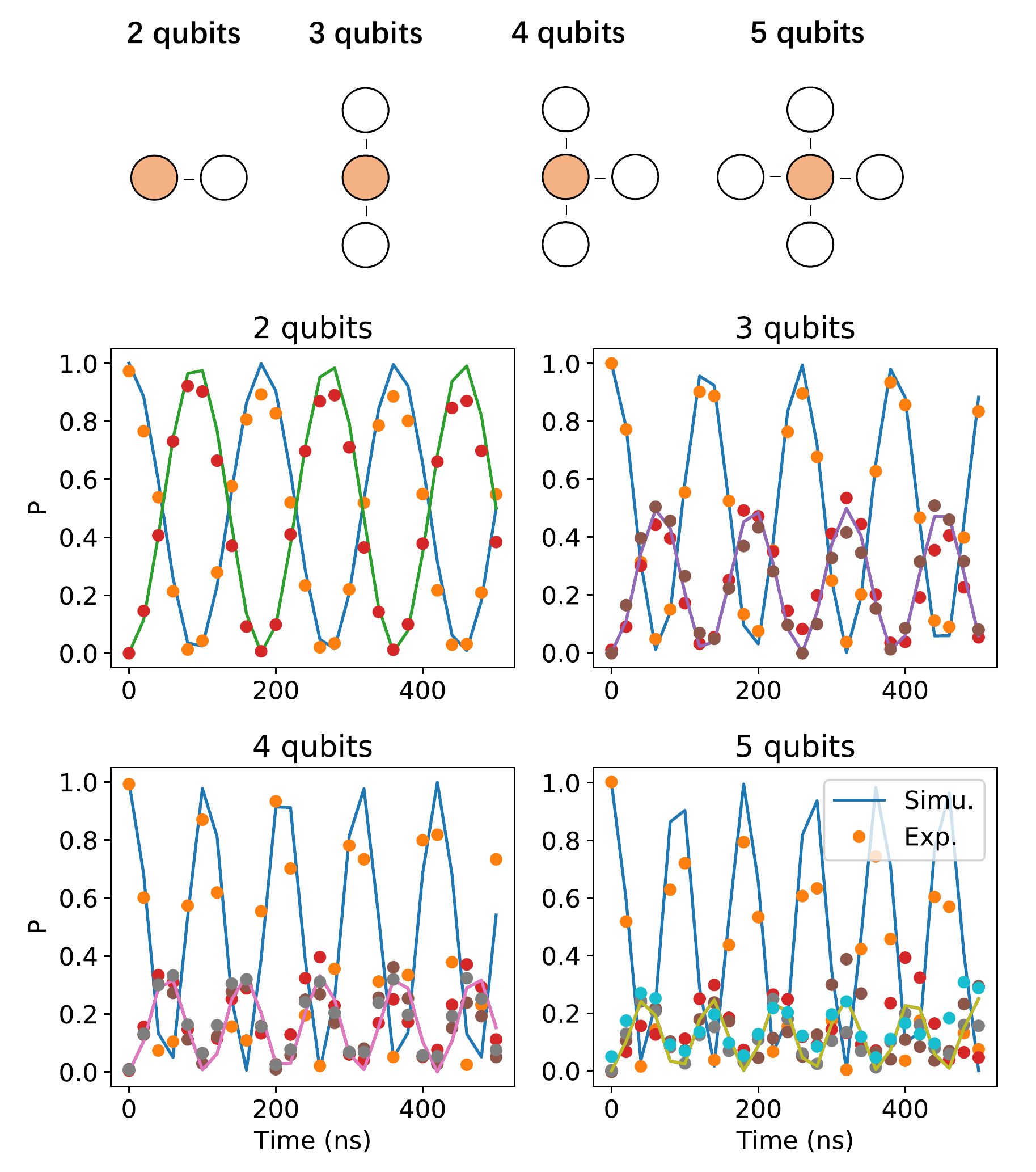}
\caption{\textbf{The coherent oscillation for different number of qubits from 2 to 5 qubits.} In the experiment, the qubit marked in orange is excited, then all qubits are detuned to the same interaction frequency, and the population of all involved qubits are measured. The solid lines correspond to numerical simulations with no disorders and the dots correspond to the experimental data. We fit the experimental data with simulations with disorder considered to extract the disorders. In these examples, the disorders are mostly negligible.}
\label{qqswap}
\end{figure*}

\blue{The Z pulse distortion correction is to remove the unwanted distortion of the control pulse applied on the flux bias of the qubit. The goal of the correction is to obtain a constant frequency $d_i$ for the time-independent evolution of the system. We utilized the calibration and correction discussed in Ref.~\cite{yan2019strongly}. We show the results before and after the correction in Fig.\ref{pulsecorrection}. After correcting the Z pulse distortion, we utilized the calibration of the qubit-frequency alignment with multi-qubit swapping at different qubit sites as discussed in Ref.~\cite{Gong2021}. Examples of the calibrations are shown in Fig.~\ref{qqswap}, where disorders have been corrected, and coherent oscillations can be observed for at least 500 ns.}

There are two qubit frequencies in this experiment: the idle points and the interacting points. The qubits are biased at their idle points to perform the single-qubit and readout operations. For each qubit, the idle points are designed to be away from their neighbors to avoid the unnecessary microwave crosstalk. Meanwhile, we optimize the idle points to maximize the single-qubit operation fidelity and readout performances following the procedures listed in Ref.~\cite{Gong2021}. The interacting point is set at 5.192 GHz for 3 by 3 instances, and 5.09 GHz for the 4 by 4 to 8 by 8 instances, respectively. Using the interacting point as the reference, we detune the qubits to the corresponding frequencies to utilize disorders. At the interacting point, the average effective coupling strength for all nearest-neighboring qubit pairs is 2.185 MHz for 3 by 3 instances, and 2.75 MHz for 4 by 4 to 8 by 8 instances. 

\begin{figure*}[hbt]
\centering \includegraphics[width=1\linewidth]{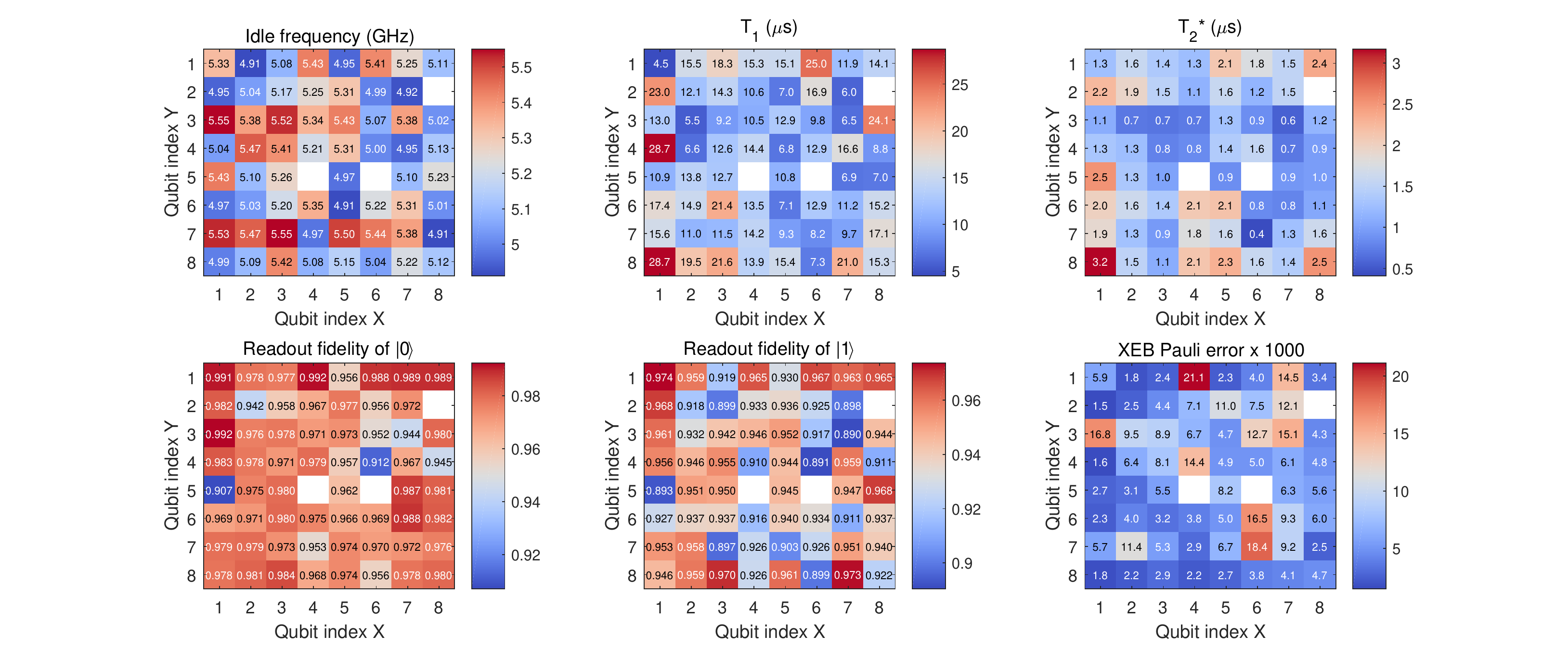}
\caption{\textbf{System parameters for our quantum processor.}
}
\label{figS1}
\end{figure*}

For state readout, we utilize the second-excited state readout technology \cite{Wang2021} to reduce the impact of state relaxation. The readout fidelity is 0.971 (0.937) for $\ket{0}$ ($\ket{1}$) state for all qubits in average. We benchmark the single-qubit gate error using cross-entropy benchmarking (XEB)~\cite{arute2019quantum}. The average XEB Pauli error is 0.65$\%$, among which 0.47$\%$ is the speckle purity benchmarking (SPB)~\cite{Aaronson2017} error and 0.18$\%$ is the control error. We summarize the performances of the quantum processor in Tab.~\ref{tableSa} and Fig.~\ref{figS1}.

\begin{figure*}[hbt]
\centering \includegraphics[width=0.5\linewidth]{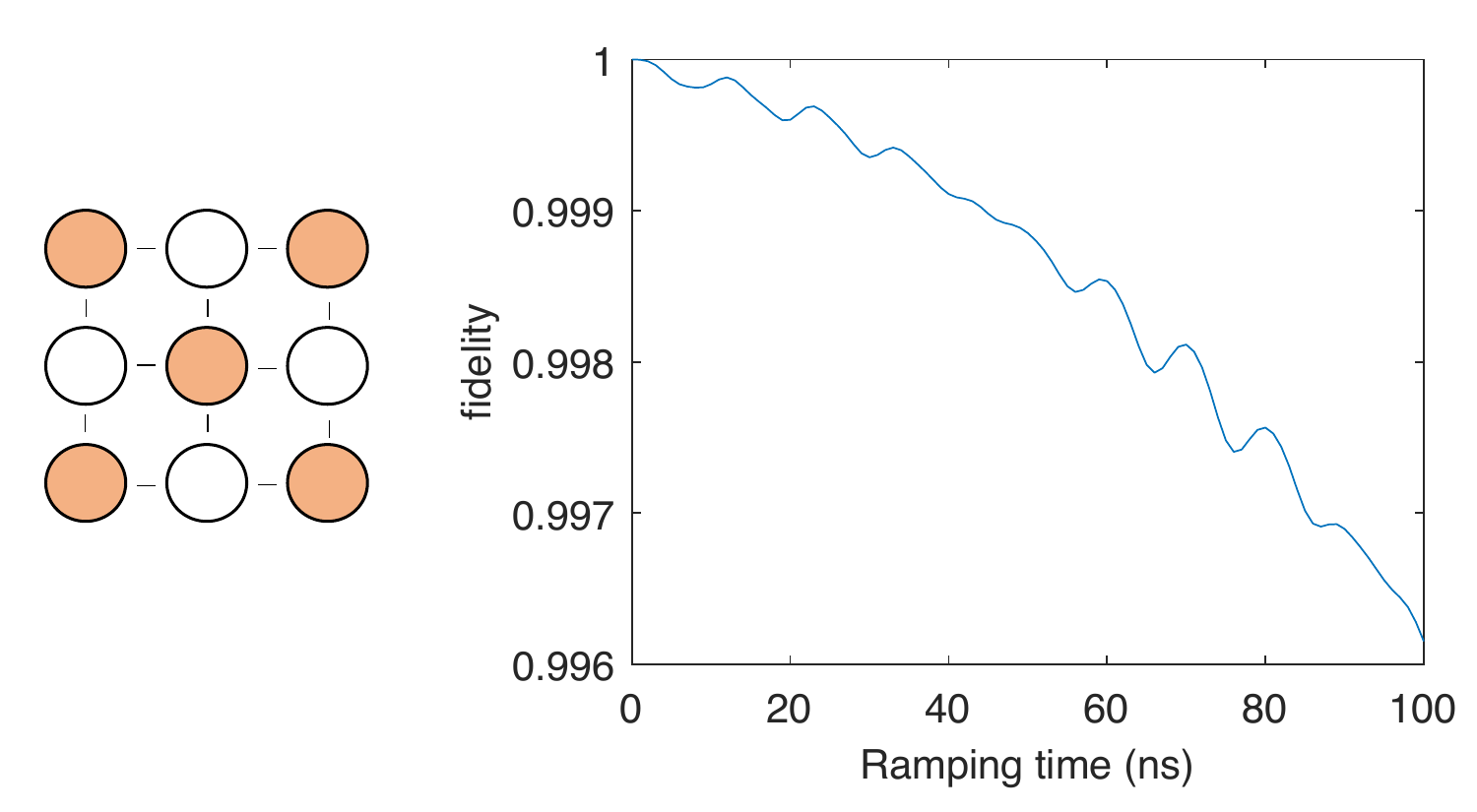}
\caption{\textbf{The simulated fidelity as a function of the ramping up and down time.} We use the squared statistical overlap as a quantification of fidelity. The system contains 9 qubits arranged in 3 by 3, with coupling strength defined as 2 MHz. The evolution time for the constant amplitude stage is 200 ns. 
}
\label{rampingtime}
\end{figure*}

\blue{Now let’s discuss the effect of ramping up and ramping down time. In our experiment, we need to tune all qubits to the chosen detuning or interacting frequency for state preparation and quantum neural network, respectively. In the tuning operation, there are a ramping up stage, a constant amplitude stage, and a ramping down stage. The ramping up and down time are extremely short in comparing with the coupling strength. In our experiment, the ramping up and down time is only 4 ns. We numerically simulated the process with 0 to 100 ns as the ramping up and down time. As shown in Fig.~\ref{rampingtime}, we use the squared statistical overlap as a quantification of fidelity defined by $F=(\sum\sqrt{p_{(i,j)}q_{(i,j)}})^2/\sum p_{i,j}\sum q_{i,j}$, and find that the fidelity keeps in a high level for the ramping time short than 10 ns. The ramping up and down in such a short time mostly contribute to the single qubit phase. In our experiment, in the state preparation, such local single qubit phase will not affect the property of the state whether it is an ergodic or localized state. In the quantum neural network, such single qubit phase can also be regarded as part of the parameters of the QNN, thus will not introduce error. }

\begin{table*}[htb]
\centering
\begin{tabular}{cccc}
\hline
Parameters& Median& Mean& Stdev.\\
\hline
Qubit idle frequency (GHz)& 5.204& 5.202& 0.196\\
$T_1$ at idle frequency ($\mu$s)& 12.94& 13.50& 5.54\\
$T_1$ at working point ($\mu$s)& 11.05& 12.13& 5.55\\
$T_2^*$ at idle frequency ($\mu$s)& 1.34& 1.41& 0.56\\
Effective coupling strength $g/2\pi$ between neighboring qubits (MHz)& 2.174& 2.185& 0.070\\
\hline
Readout fidelity of $\ket{0}$ $f_{00}$& 0.975& 0.971& 0.016 \\
Readout fidelity of $\ket{1}$ $f_{11}$& 0.940& 0.937& 0.023 \\
Effective qubit temperature (mK)& 65& 66& 11\\
\hline
Single-qubit SPB Pauli error& 0.0042& 0.0047& 0.0029\\
Single-qubit XEB Pauli error& 0.0050& 0.0065& 0.0046\\
Single-qubit control error& 0.0007& 0.0018& 0.0026\\
\hline
\end{tabular}
\caption{\textbf{Performance parameter of the ``Zuchongzhi 1.0'' quantum processor.}}
\label{tableSa}
\end{table*}

\section{level statistics}
The quantum system can be in distinct regimes~\cite{Kuhn_2007,Escalante2018,Zhang2021,abanin2019colloquium}, ergodic and localized regimes, depending on the disorder strength $h/g$. The system thermalizes under the dynamics and follows the eigenstate thermalization hypothesis~\cite{mori2018thermalization} for low disorder strengths, and would be in the localized phase for large disorder strengths~\cite{Kuhn_2007,Escalante2018,Zhang2021,abanin2019colloquium}.

Level statistics is one of the standard approach to distinguish the different phases~\cite{Manai2015,White2020,sierant2019level,sierant2020model,atas2013distribution}. The gap ratio of consecutive spacings between energy levels

\begin{equation}
{r_n} = \min \{ {\delta _n},{\delta _{n - 1}}\} /\max\{ {\delta _n},{\delta _{n - 1}}\}
\end{equation}
is introduced as a simple probe of the level statistics in Ref.~\cite{oganesyan2007localization}, where ${\delta _n} = {E_{n + 1}} - {E_n}$ is an energy difference between two consequtive levels. The average gap ratio, $\bar r$, is different for ergodic systems ($\bar r \approx 0.527$) and for localized systems ($\bar r \approx 0.386$), as analyzed in Ref.~\cite{atas2013distribution}.

\begin{figure*}[hbt]
\centering \includegraphics[width=0.7\linewidth]{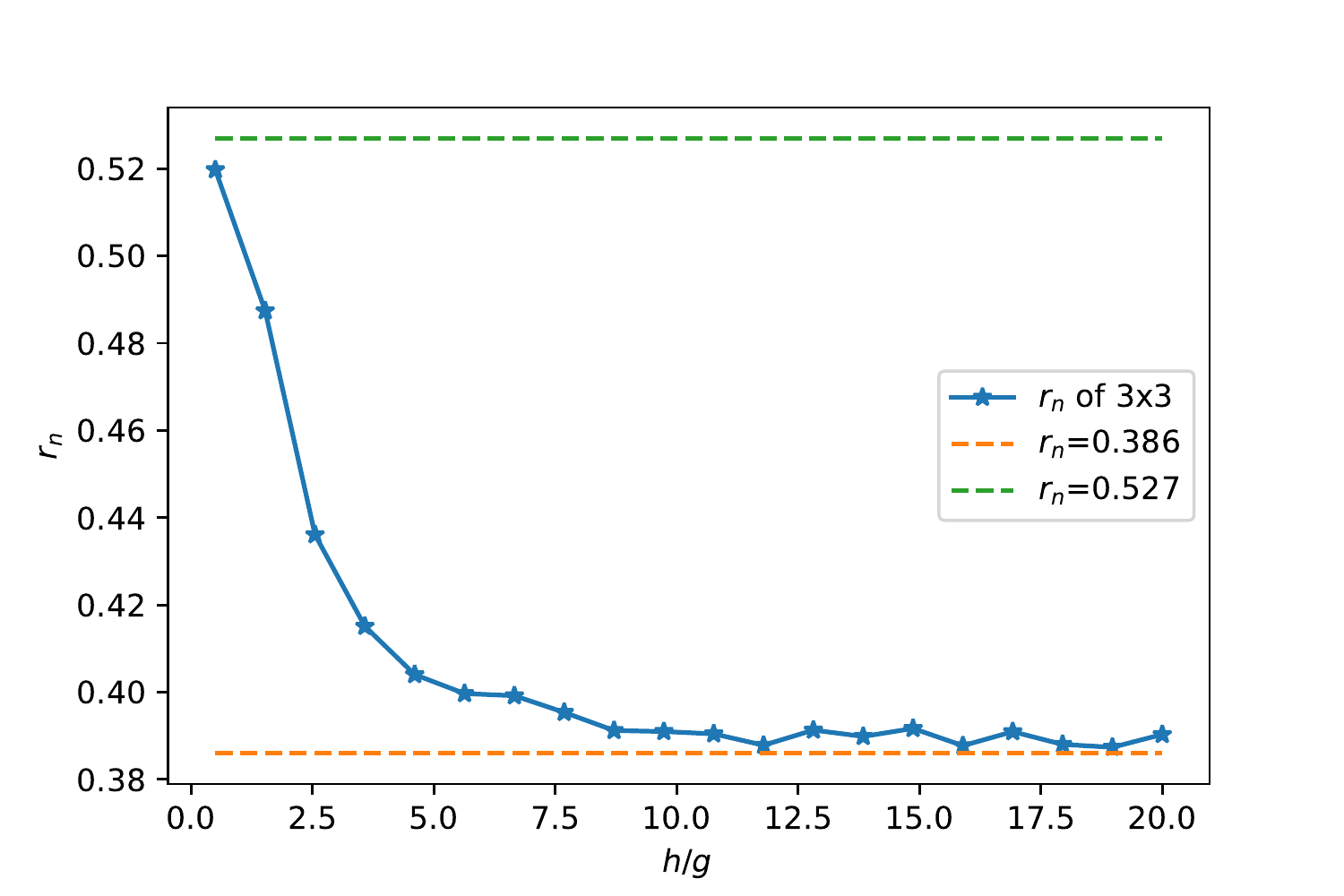}
\caption{\textbf{Level statistics of the $3\times 3$ system.} Average level spacing $\bar r$ as a function of the disorder strength $h/g$.
}
\label{level_statistic}
\end{figure*}

We perform numerical simulation to analyze the level statistics of the $3\times3$ system. Figure~\ref{level_statistic} shows the average gap ratio $\bar r$ as a function of disorder strength $h/g$ and captures the phase transition from the ergodic to the localized phase. We note that although level statistics can be used to probe these two phases, the calculation of level statistic is very expensive for large system, since the complexity of calculating the eigenvalues of a $N$-qubit system is O($2^{3N}$), which grows exponentially over the number of qubits.

\section{Imbalance dynamics}

\begin{figure*}[hbt!]
    \centering \includegraphics[width=1\linewidth]{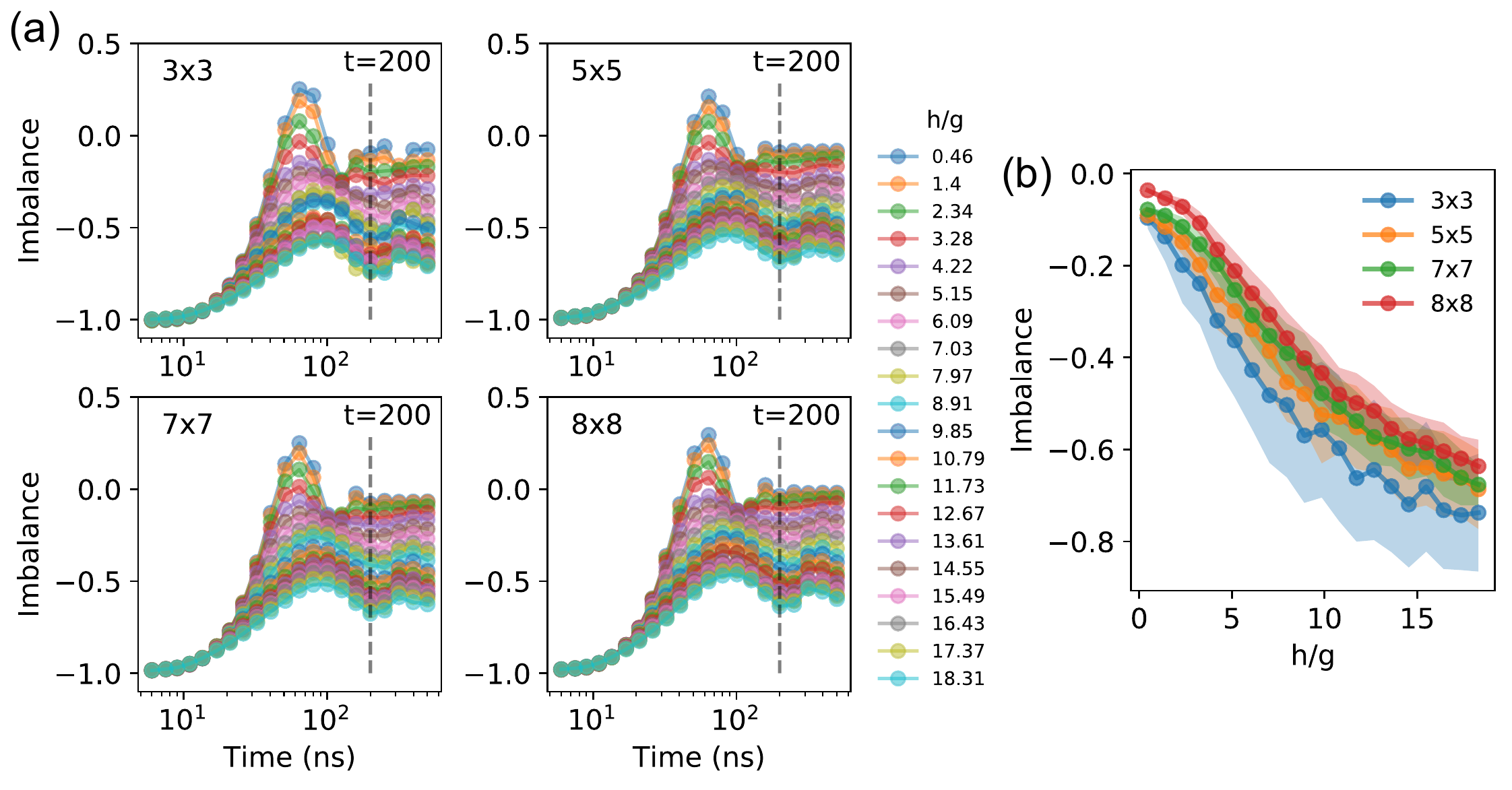}
    \caption{\textbf{Evolution of the imbalance.} (a) Time evolutions of the system imbalance at different disorder strengths and system sizes. Each data point is averaged over 50 disorder realizations. (b) The quasi-steady-state imbalance taken at 200 ns as functions of disorder strengths for different system sizes. The shadow is the error estimated via the standard deviation in the 50 disorder realizations.
    }
    \label{figS2}
    \end{figure*}

In our experiment, the ergodic states and localized states are generated by the system's evolution in the absence and presence of disorder, respectively. Here, we apply the order parameter of imbalance, defined as $\mathcal{I}=(N_e-N_o)/(N_e+N_o)$, to monitor the evolution of the system, where $N_e$ ($N_o$) is the total number of excitation quanta on the even (odd) number sites. Imbalance is an effective order parameter that reflects the preservation of the local magnetizations of the initial state. The evolution of the imbalance of the quantum system with different system sizes and different disorder strengths is depicted in Fig.~\ref{figS2}A. We found that at long times above 150~ns, the imbalance reaches a steady state. The imbalance of the steady-state approaches 0 for $h/g\approx 0$, where $h$ is the disorder strength and $g\approx 2.185~\text{MHz}$ the effective average coupling strength between neighboring qubits. However, if the disorder strength increases, the steady-state imbalance becomes larger, signaling the breakdown of the ergodicity. Meanwhile, we also provide the results of imbalance at time $t=200~\text{ns}$ as a function of the disorder strength for different system sizes (see Fig.~\ref{figS2}B). We can observe that as the system size increases, the slope of the curve becomes smaller, indicating that more disorder is required to enter the localized phase for a larger system. It is worth noting that these experimental results have gone beyond the classical simulatable area and have never been produced in theory or experiment to the best of our knowledge, which can provide good guidance for future research.

\section{Numerical simulations}

\begin{figure*}[!hbt]
\centering \includegraphics[width=0.6\linewidth]{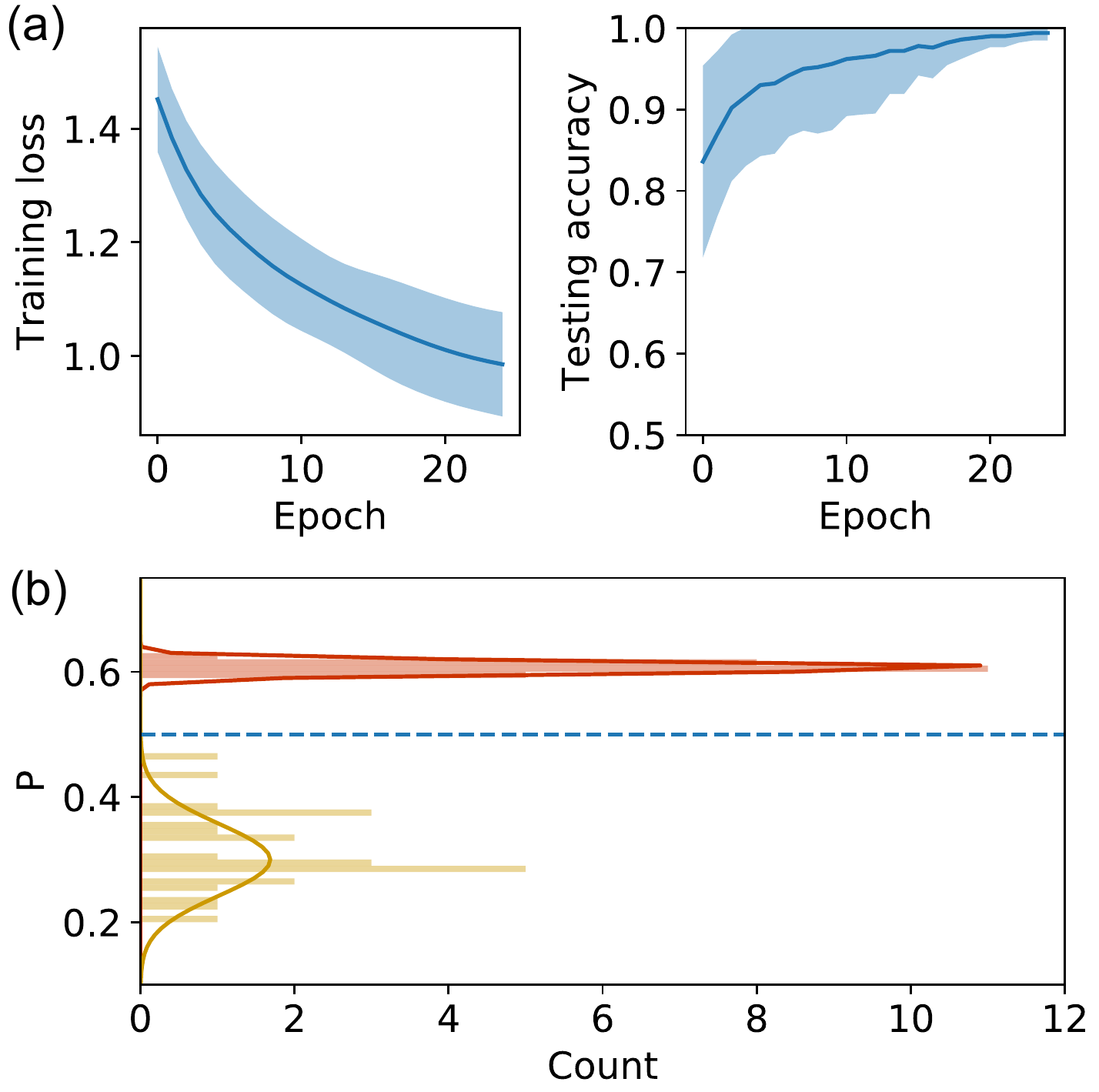}
\caption{\textbf{Numerical simulation of Quantum Neuronal Sensing for the 3x3 system size.} (a) The loss and accuracy at every epoch during the training, respectively, where each point is the average of 10 independent training instances of the QNN. (b) Results of applying the QNN at the last epochs to the testing data set containing 50 quantum states. The two different groups are well separated after 25 epochs of training. The solid lines represent the Gaussian fitting to the distributions of data.
}
\label{3_3_simulation}
\end{figure*}

Here we provide the numerical simulation results of applying the quantum neuronal sensing to the small-scale $3\times3$ quantum system (see Fig.~\ref{3_3_simulation}). The setting of the parameters in our simulation is consistent with the experiment. The training (testing) data has 20 (50) individual states, respectively, of which half are ergodic states and the remaining localized states. In Fig.~\ref{3_3_simulation}A,B we plot the loss and accuracy values of 25 epochs during the training of the QNN, where each point is the average of 10 independent training instances of the QNN with initial parameters and dataset randomly generated. In Fig.~\ref{3_3_simulation}C, we show the measurement results of the readout qubit when the QNN of the last training epochs is applied to the testing data. The testing data are clearly distinguished into two groups at the end of the last epoch. By comparing with Figure 2 in the main text, we can find that the experiment is in good agreement with the numerical simulation, indicating the high quality of our experiment.

\hhl{
Besides, we note that the QNN with single-qubit measurement is sufficient for the classification task on $3\times3$ quantum systems. This brings great convenience to the scalability of QNN, mainly due to the following two factors:

1)	As proven in Ref.~\cite{cerezo2021cost}, cost function with global observables leads to exponentially vanishing gradients (i.e., barren plateaus), while cost function with local observables leads to at worst a polynomially vanishing gradient.

2)	Experimentally, the noise introduced by single-qubit measurement is obviously much smaller multi-qubit measurements, and single-qubit measurements are immune to correlated measurement noise.

Since we need to scale the system to 61 qubits, we chose local observables to alleviate the problems of barren plateau and measurement noise as it scales up. Fortunately, the results of our large-scale experiments (see Fig.3 in the main text) show that our employed architecture works for the given problem, indicating that this architecture has the expressive power to solve this specific problem.}

\section{Extended data}

In our experiment, the evolution time of the analog circuit part in the QNN is set to 200~ns. These settings can also be regarded as hyperparameters of our QNN.

\subsection{The hyperparameter of evolution time in the analog circuit part}

\begin{figure*}[!hbt]
\centering \includegraphics[width=0.6\linewidth]{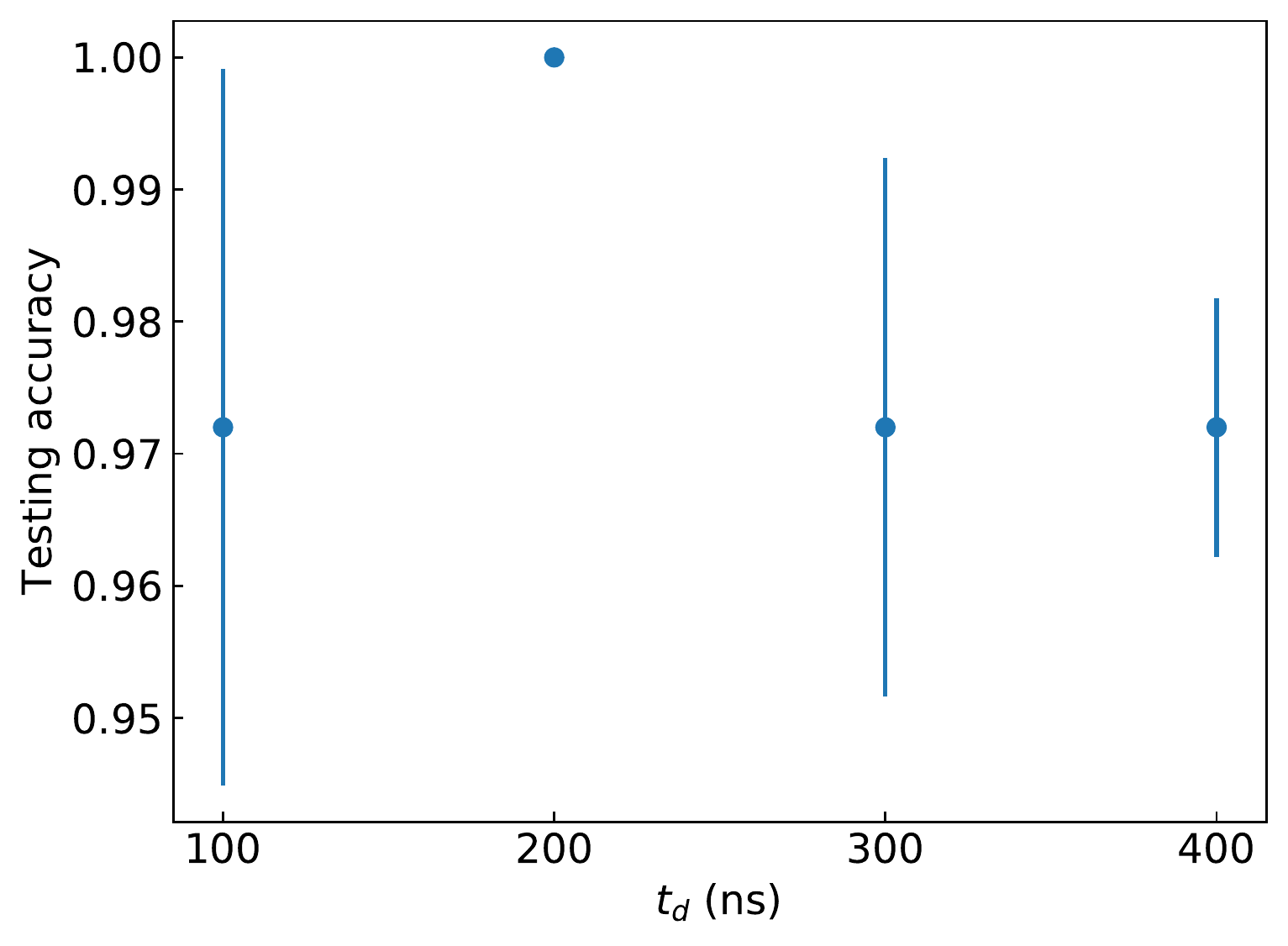}
\caption{\textbf{Classification accuracy of the QNN with different evolution times $t_d$ in the analog circuit part.} For the $3\times 3$ system, the classification accuracy of the trained QNN model for various evolution times $t_d=100~\text{ns}, 200~\text{ns}, 300~\text{ns},~\text{and}~400~\text{ns}$ used in the analog circuit part on the testing data, where each point is the average of 5 individual trainings of the QNN.
}
\label{fig_td}
\end{figure*}

We also experimentally tested the performance of the QNNs with different evolution times $t_d$ in the analog circuit part. Figure~\ref{fig_td} shows the classification accuracy of the QNNs for various times $t_d=100~\text{ns}, 200~\text{ns}, 300~\text{ns},~\text{and}~400~\text{ns}$ used in the analog circuit part, respectively. In all four cases, the trained models can achieve high classification accuracy on the testing data. And the stability of the results and the classification accuracy are relatively higher when $t_d = 200~\text{ns}$. Thus, we finally set the default evolution time of the analog circuit part in the QNN as $t_d=200~\text{ns}$.

\begin{figure*}[!hbt]
\centering \includegraphics[width=0.6\linewidth]{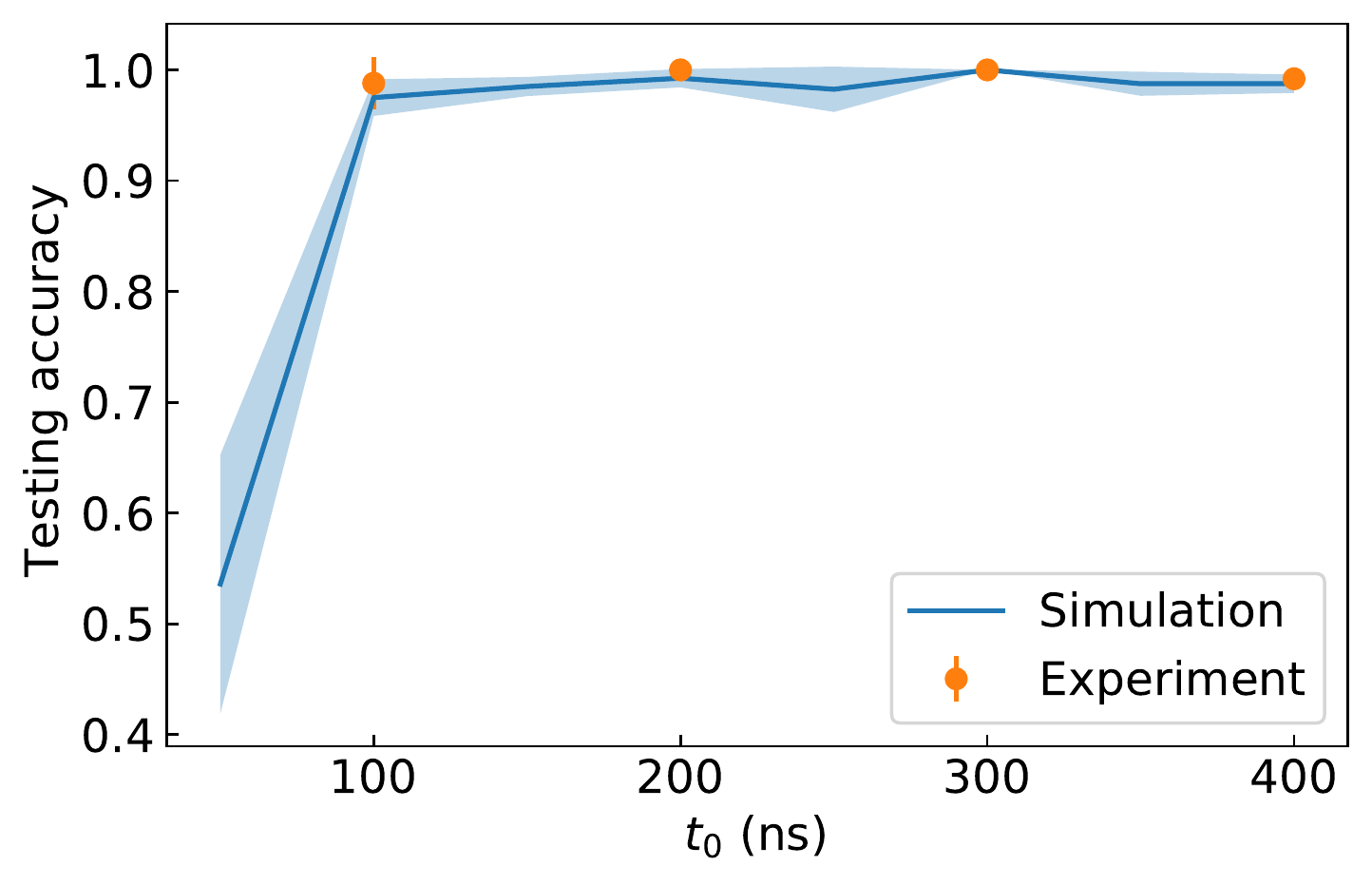}
\caption{\textbf{Classification accuracy of the QNN at different evolution time $t_0$.} Presented are the classification accuracy of the QNN on the testing data for the $3\times 3$ system at different evolution time $t_0=100~\text{ns}, 200~\text{ns}, 300~\text{ns},~\text{and}~400~\text{ns}$, where each point is the average of 5 individual trainings of the QNN. \blue{The solid line and the shadow correspond to the numerical simulations.}
}
\label{fig_t0}
\end{figure*}

\subsection{Classification of quantum states at other evolution times} As analyzed in Section 2, at long times around 200 ns or above, the system reaches a steady-state. In our experiments, after the system has evolved for 200 ns, we then distinguish whether the system's state at this time is ergodic or localized. As a general quantum neuronal sensing, our approach, in principle, can be used to distinguish the quantum states with different properties. Figure~\ref{fig_t0} shows the results of applying the quantum neuronal sensing to distinguish the quantum states generated from the system in the absence and presence of disorder, after system has evolved for different times $t_0=100~\text{ns}, 200~\text{ns}, 300~\text{ns},~\text{and}~400~\text{ns}$. For quantum states at different evolutionary moments, our approach can distinguish the quantum state from a system with or without a disorder with a high success rate, showing the universality and scalability of our approach.

\begin{figure*}
    \centering \includegraphics[width=0.8\linewidth]{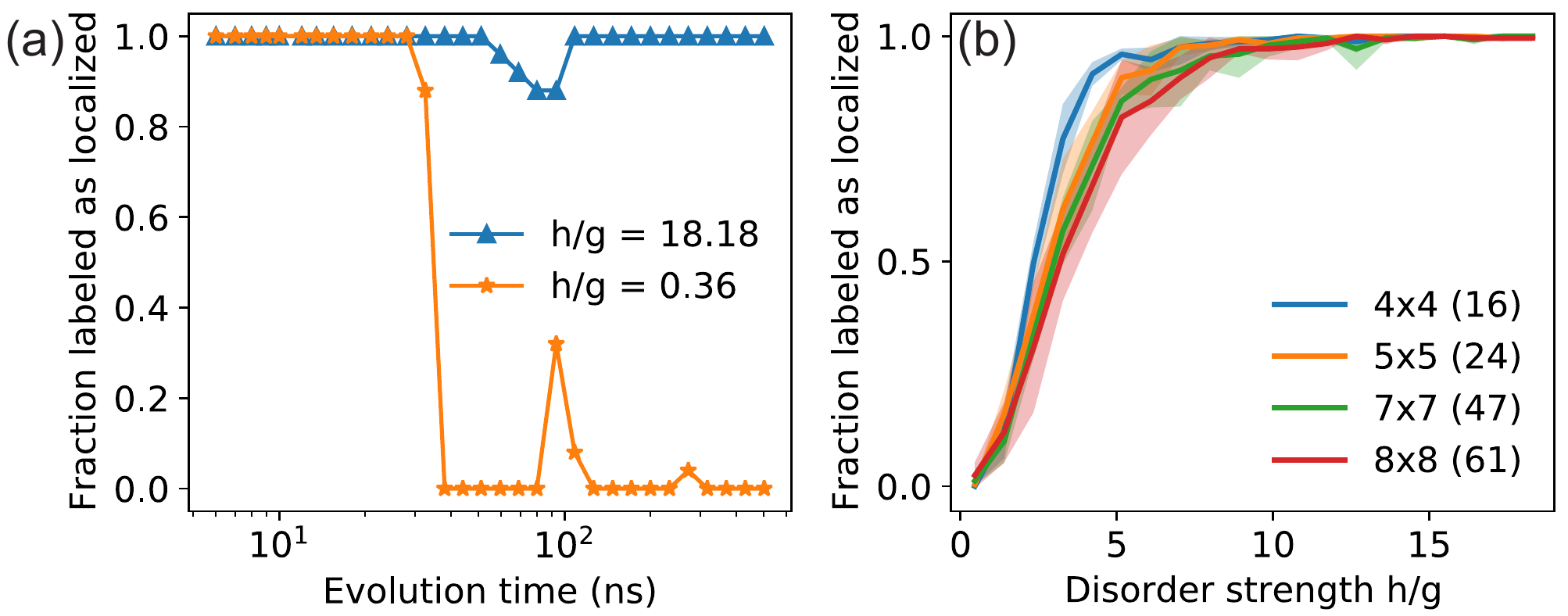}
    \caption{\hhl{\textbf{The results of applying the trained QNN to the quantum states generated by different evolution time and disorder strengths.} (a) Statistical results for the quantum states generated with different Hamiltonian evolution time and two different disorder strengths for $3\times3$ system. After training the QNN using the quantum states generated with 200 ns evolution time, the trained QNN is applied to monitor the dynamical evolution of two types of quantum states. The evolution time for state preparation is arranged logarithmically between 6 and 501 ns. At each time point, 50 sets of parameters in the testing set, of which $h/g=0.36$ and $18.18$ each account for half, are used to generate the two classes of states. In this series of experiments, the initialization strategy is used before training the QNN, to achieve better performance in a small number of training epochs. Besides, we optimize the threshold for separation as 0.47 (the intersection of the Gaussian fitting of the classification results of the two classes in the training set), as the classification results are somewhat imbalanced. (b) The probability of quantum states classified in localization for the states generated with a variety of disorder strength. For each disorder strength, we randomly generated 50 disorder profiles and the Hamiltonian evolution time for state preparation is 200 ns. $h/g$ are equally spaced ranging from 0.46 to 18.3. }}
    \label{fig4}
    \end{figure*}
    
    \begin{figure*}[hbt!]
        \centering \includegraphics[width=0.6\linewidth]{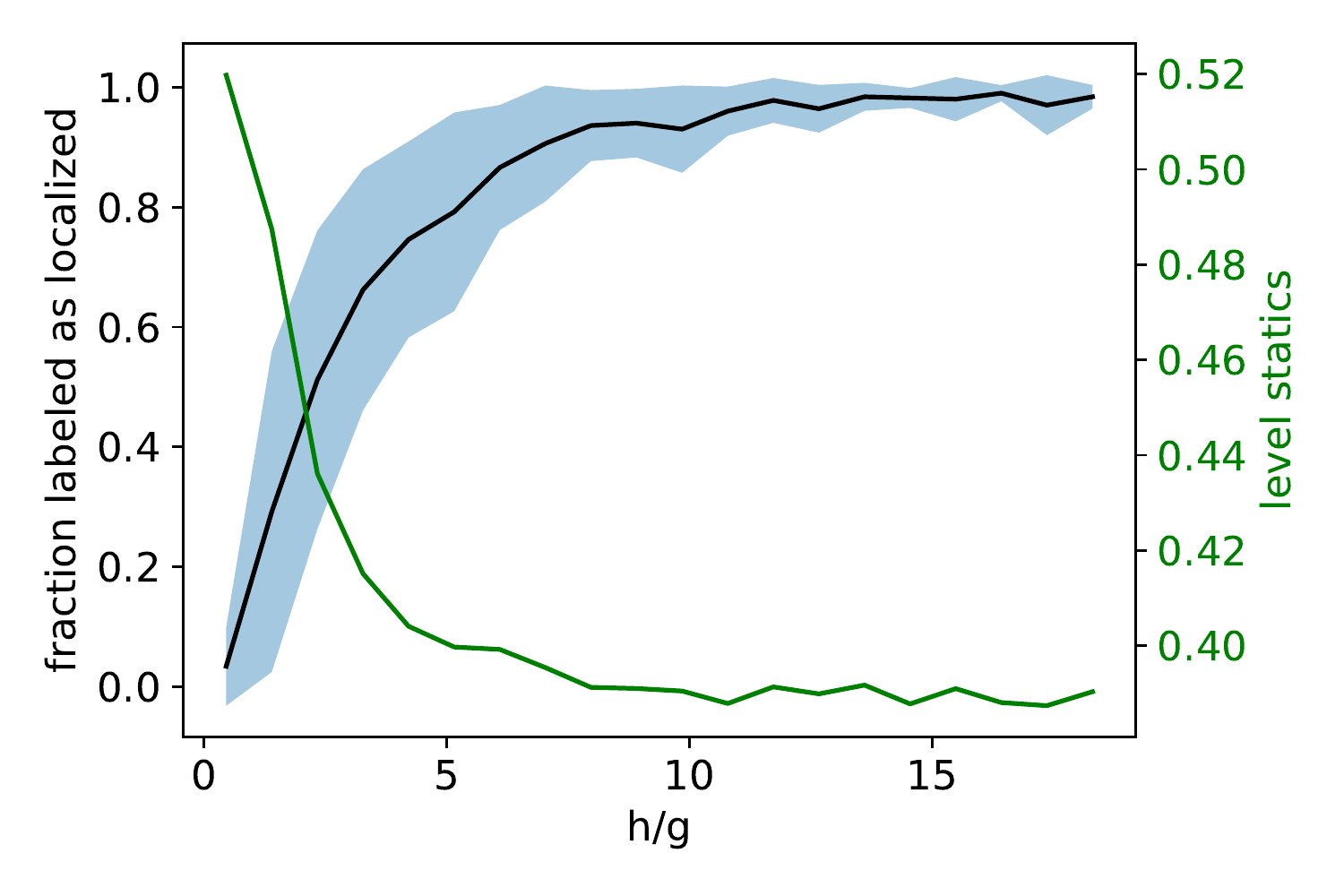}
        \caption{\hhl{\textbf{The experimental results of classification with trained QNN and the numerical results of level statistics for the $3\times 3$ system with various disorder strengths.} The black solid line is the experimental probabilities of quantum states classified in localization by the trained QNN. The blue shadow is the standard deviation of the results from 10 individual instances. The green solid line is the numerical level statistics. These two curves show a strong correlation.}}
        \label{fig_predvsls}
        \end{figure*}

\subsection{\hhl{Generalization of the trained QNN}}

\hhl{The above results demonstrate that our approach has the ability to discriminate each quantum state produced by the given two types of systems with high accuracy. We further investigate more potential applications of the trained QNN by applying it to identify the statistics of quantum systems with different settings than the training set. In Fig.\ref{fig4}\textbf(a), we show that the trained QNN can be used to monitor the dynamical evolution of quantum states with large and small disorder strength, and statistically the localized and ergodic states can be clearly distinguished by QNN when the evolution time is higher than about 40 ns. In Fig.\ref{fig4}\textbf(b), we show that the trained QNN can be used to predict the probability of the localized quantum states with different disorders. We select 20 values equally spaced in between $h_{erg}$ and $h_{loc}$. For each selected $h$, we randomly generate 50 sets of disorders uniformly distributed in $[-h, h]$. We count the probability that the quantum state in each data set being classified as a localized state. This probability reflects the closeness of the quantum states generated under a given disorder to the trained localized state. Then, with the well-trained QNN for different system size, we also predicted the probability of the localized quantum states with different disorders. As shown in Fig.\ref{fig4}\textbf(b), it is observed that the probability of the quantum states in localization increases together with the increment of the disorder as expected, which shows highly consistent with the prediction by level statistics~\cite{luitz2015many,sierant2020model} (see Section D for more discussion), suggesting that our protocol can provide us a different perspective to the study of quantum phases such as ergodic and localized states. 
These observations clearly demonstrate the strong generalization of QNN, providing us a positive signal that our quantum enhanced processing technique might have a wide range of applications in the study of many-body physics.}

\hhl{
\subsection{\hhl{Correlation between QNN and level statistics}}

We have shown in Fig.~\ref{fig4}(b) that the trained QNN can be utilized to predict the probability of the localized quantum states with different disorders. The level statistics can be also utilized for comparable tasks, as shown in section II. Figure~\ref{fig_predvsls} shows the experimental result of trained QNN and the numerical result of level statistics for the $3\times 3$ system with various disorders. We calculate the correlation coefficient $R$ of the covariance matrix of these two arrays of results in Fig~\ref{fig_predvsls}. The correlation coefficient $R$ of two arrays is defined as

\[R = \frac{{{C_{12}}}}{{\sqrt {{C_{11}}{C_{22}}} }}\]
where $C$ is the covariance matrix of two arrays. The value of $R$ range from -1 to +1, and the larger the  absolute value of $R$, the larger is the correlation value. If the correlation coefficient $R$  is less than 0 then it negatively correlated. And if it’s greater than 0 then it’s positively correlated. The calculated correlation between the two arrays of results is $-0.936 \pm0.044$. It shows that these two arrays of results are highly negatively correlated, indicating the trained QNN has the potential to be an order parameter like the level statistics for many-body quantum physics. Moreover, the computational cost of level statistic grows exponentially over the number of qubits, as shown in section II. However, our QNN is very efficient, it only requires a shallow parametered quantum circuit and only single-qubit measurement.
}

\begin{figure*}[hbt!]
    \centering \includegraphics[width=0.8\linewidth]{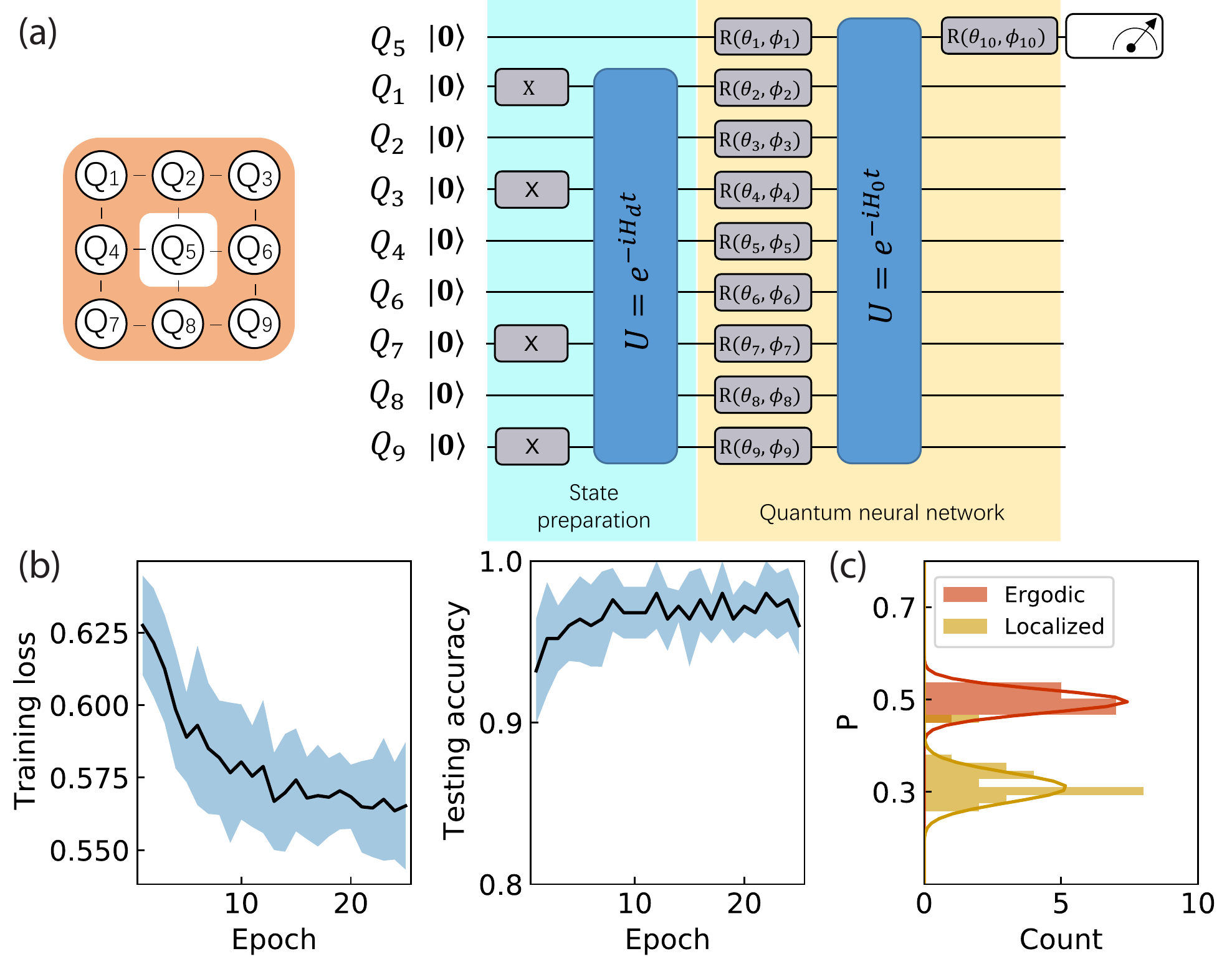}
    \caption{\hhl{\textbf{Classification of quantum states using an ancilla qubit.} (a) The quantum circuit of the protocol that using Q5 as a probe qubit to learn the properties of the quantum system. (b) The loss and accuracy at every epoch during the training. (c) The results of applying the QNN at the last epochs to the testing data set containing 50 quantum states.}}
    \label{fig9}
    \end{figure*}

\hhl{
\subsection{\hhl{Classification of quantum states using an ancilla qubit}}

We particularly designed an experiment to further demonstrate that our quantum neuronal sensing approach can be utilized as a probe to learn the properties of the quantum system. In this experiment, the qubit measured, which is called  ancilla or probe qubit, after running the quantum neural network were not among those involved in the many-body quantum state. Specifically, our experiment uses the middle qubit, marked as Q5 in Fig.~\ref{fig9}(a) in the $3\times3$ grid for this role, and the 8 outer qubits for preparing the many-body state. The equivalent quantum circuit is shown on the right side of Fig.~\ref{fig9}(a). The experimental results are shown in Fig.~\ref{fig9}(b,c). During the training procedure, the parameter initialization strategy introduced in the main text is employed to determine a good initial parameters.  As can be seen from the curves of loss and accuracy (see Fig.~\ref{fig9}(b)), the training procedure can further improve the performance of QNN. And finally, the two classes of ergodic and localized states can be distinguished by detecting the probe qubit (see Fig.~\ref{fig9}(c)). After the QNN, there appears to be a minor class imbalance; however, by simply adjusting the threshold to around 0.41, we can get classification accuracy as high as 0.960$\pm$0.018, which is averaged over five instances.

We note that the probe qubit is completely unaware of the quantum system's state prior to QNN processing because it is not involved in the quantum system's evolution. However, following QNN processing, the probe qubit can be used to distinguish whether the system is local or ergodic. Such an experiment shows clearly that the distinguishability could credibly be attributed to a non-trivial behavior of the QNN.}
\bibliography{references}